\documentclass[aps,prd,twocolumn,nofootinbib,superscriptaddress]{revtex4-1}
\pdfoutput=1
\usepackage[T1]{fontenc}
\usepackage[utf8]{inputenc}
\usepackage{lmodern, amsmath, amssymb, subcaption, epsfig, cancel}
\usepackage[colorlinks=true, citecolor=blue, urlcolor=blue, linkcolor=blue, breaklinks=true, bookmarksopen]{hyperref}

\allowdisplaybreaks[3]

\def\equationautorefname~#1\null{Eq.\,(#1)\null}
\def\pageautorefname\nobreakspace{p.}

\usepackage{ifpdf,feynmp}
\ifpdf\fi
\AtBeginDocument{\begin{fmffile}{graph}
\fmfcmd{%
prologues:=3;
thin := .5pt;
arrow_ang := 20;
arrow_len := 3.5thick;
curly_len := 2.5thick;
color mb, mr, mg; mb :=.5blue; mr :=.5red; mg :=.5green;
style_def top expr p =
  save oldpen;
  pen oldpen;
  oldpen := currentpen;
  pickup oldpen scaled 3.5;
  ccutdraw p;
  pickup oldpen scaled 1.5;
  cullit; undraw p; cullit;
  cfill (arrow p);
enddef;
}
}
\AtEndDocument{\end{fmffile}}

\newcommand{\ie}{{\it i.e.~}}

\newcommand{\lag}{{\cal L}}

\DeclareMathOperator{\D}{d\!}

\newcommand{\hc}{\ensuremath{\mathrm{h.c.}}}
\renewcommand{\lag}[1]{\ensuremath{\mathcal{L}_\mathrm{#1}}}

\newcommand{\FDF}[1][]{\varphi^\dagger\!#1\!\overleftrightarrow{D}\!_\mu\:\varphi}
\newcommand{\FDFI}[1][]{\varphi^\dagger\!#1\!\overleftrightarrow{D}^I\!\!\!_\mu\:\varphi}

\let\Re\undefined \DeclareMathOperator{\Re}{Re}
 
\DeclareMathOperator{\Br}{Br}

\DeclareMathAlphabet{\mathsfit}{\encodingdefault}{\sfdefault}{m}{sl}
\newcommand{\ges}[1]{\mathsfit{#1}}
\newcommand{\ind}[2]{^{#1}_{\hphantom{#1}#2}}

\makeatletter
\newcommand{\parenbar}{\mathpalette\p@renb@r}
\def\p@renb@r#1#2{%
	\vbox{%
		\ifx#1\scriptscriptstyle\dimen@.7em\dimen@ii.20em\else%
		\ifx#1\scriptstyle	\dimen@.8em\dimen@ii.25em\else%
					\dimen@1.em\dimen@ii.4em\fi\fi%
		\offinterlineskip%
		\ialign{%
			\hfill##\hfill\cr
			\vbox{\hrule width\dimen@ii}\cr
			\noalign{\vskip-.3ex}%
			\hbox to\dimen@{$\mathchar300\hfil\mathchar301$}\cr
			\noalign{\vskip-.3ex}%
			$#1#2$\cr%
		}%
	}%
}%
\newcommand{\us}[2]{\underset{\ifx#2\@empty\else#2\%\fi}{#1}}
\makeatother
\newcommand{\usv}[2]{\underset{\text{---}}{#1}}
\newcommand{\n}{2mm}

\begin{document}

\rightline{CP3-14-85}
\title{A global approach to top-quark flavor-changing interactions}
\newcommand{\louvain}{Centre for Cosmology, Particle Physics and Phenomenology, 
Universit\'e catholique de Louvain, B-1348 Louvain-la-Neuve, Belgium}
\author{Gauthier Durieux}
\affiliation{Laboratory for Elementary Particle Physics, Cornell University, Ithaca, NY 14853, USA}
\affiliation{\louvain}
\author{Fabio Maltoni}
\affiliation{\louvain}
\author{Cen Zhang}
\affiliation{Department of Physics, Brookhaven National Laboratory, Upton, NY, 11973, USA}

\begin{abstract}
We adopt a fully gauge-invariant effective-field-theory approach for
parametrizing top-quark flavor-changing-neutral-current interactions. It 
allows for a global interpretation of experimental constraints (or measurements)
and the systematic treatment of higher-order quantum corrections. We discuss
some recent results obtained at next-to-leading-order accuracy in QCD and
perform, at that order, a first global analysis of a subset of the available
experimental limits in terms of effective operator coefficients. We encourage
experimental collaborations to adopt this approach and extend the analysis  
by using all information they have prime access to.
\end{abstract} 


\maketitle

\section{Introduction}

The wealth of top quarks produced at the LHC has moved top physics to a
precision era. Detailed information on the top couplings, their strengths as well
as Lorentz structures, has been collected and possible deviations are being
constrained. In addition, interactions that are absent or suppressed in the
standard model (SM) become more and more accessible. Among these, top-quark
flavor-changing-neutral-current interactions (FCNCs) play a special role. Highly
suppressed by the Glashow-Iliopoulos-Maiani mechanism, the SM predicts them to
be negligible. Branching ratios for top FCNC decays are notably of the order of
$10^{-12}- 10^{-15}$~\cite{Mele:1998ag, Eilam:1990zc, AguilarSaavedra:2002ns} in
the SM. Any evidence for such processes would thus immediately point to new
physics. In addition, the recent discovery of a scalar particle closely
resembling the SM Higgs boson~\cite{Aad:2012tfa, Chatrchyan:2012ufa} has made
Higgs-mediated FCNCs experimentally searchable.

A wide variety of limits have been set on top-quark FCNC interactions, see, {\it
e.g.}, Ref.~\cite{Agashe:2014kda}. Single-top $p\parenbar{p}\to t$ production has
been searched at the Tevatron by CDF~\cite{Aaltonen:2008qr} and at the LHC by
ATLAS \cite{TheATLAScollaboration:2013vha, Aad:2012gd} while
D0~\cite{Abazov:2010qk, Abazov:2007ev} and CMS~\cite{CMS:2014ffa} considered the
$p\parenbar{p}\to t j$ production mode.
In addition, CMS also searched for single-top production in association with a
photon~\cite{CMS:2014hwa} or a charged lepton pair~\cite{CMS:2013nea}.
At LEP2, $e^+e^- \to t\,j$ has been investigated by all four
groups~\cite{LEP2001, Abbiendi:2001wk, Heister:2002xv, Achard:2002vv,
Abdallah:2003wf,DELPHI:2011ab} while, at HERA, the single-top $e^-\,p\to e^-t$ production
has been considered by ZEUS~\cite{Abramowicz:2011tv, Chekanov:2003yt} and
H1~\cite{H1, Aaron:2009vv, Aktas:2003yd}.
The FCNC decay processes, $t\to j\,\ell^+\ell^- $ and $t\to j\,\gamma$,
have also been studied, at the Tevatron by CDF \cite{Abe:1997fz,
Aaltonen:2008ac, Aaltonen:2009ef} and D0 \cite{Abazov:2011qf}, and at the LHC by
ATLAS \cite{Aad:2012ij, ATLAS:2011mka, ATLAS:2011jga} and CMS
\cite{Chatrchyan:2013nwa, Chatrchyan:2012hqa}.
Finally, $t\to j\,h$ has been constrained by CMS~\cite{CMS:2014qxa} that
combined the leptonic $WW^*$, $\tau\tau$, $ZZ^*$ and $\gamma\gamma$ channels
while ATLAS used the last (and most sensitive) one
only~\cite{TheATLAScollaboration:2013nia, Aad:2014dya}.

The effective field theory (EFT)~\cite{Weinberg:1978kz,
Weinberg:1980wa, Georgi:1994qn} is a
particularly relevant framework for parametrizing new physics and has been used in
many top-quark FCNC studies~\cite{BarShalom:1999iy,AguilarSaavedra:2000aj,
CorderoCid:2004vi, AguilarSaavedra:2004wm, Ferreira:2005dr, Ferreira:2006xe,
Ferreira:2006in, Guedes:2008hu, Ferreira:2008cj, Coimbra:2008qp, Kao:2011aa,
Atwood:2013ica, Agram:2013koa, Khatibi:2014via, Greljo:2014dka, Khanpour:2014xla}. It does not only
incorporate all possible effects of new heavy physics in a model-independent
way, but also orders them and allows us to consistently take into account
higher-order quantum corrections. Leading-order (LO) predictions are actually
insufficient when an accurate interpretation of observables in terms of theory
parameters is aimed at. QCD corrections in top-decay processes~\cite
{Kidonakis:2003sc, Zhang:2008yn, Drobnak:2010wh, Drobnak:2010by, Zhang:2010bm,
Zhang:2014rja} typically amount to approximately $10\%$, while they can reach
between $30\%$ and $80\%$ in production processes~\cite{Liu:2005dp,
Gao:2009rf, Zhang:2011gh, Li:2011ek, Wang:2012gp}. The running and mixing of
operator coefficients should also be taken into account. While an EFT
description in principle requires a complete basis of operators to be used,
neglecting some of them may appear consistent when only lowest-order estimates
of specific processes are considered. The next-to-leading-order (NLO)
counterterms as well as the renormalization-group (RG) running and mixings of
operator coefficients however clearly reveal the unnatural and inconsistent
character of neglecting some operators. A proper EFT description of new physics
should necessarily be global. Currently, however, the limits obtained by
experimental collaborations almost always assume one single FCNC interaction is
present at the time.

The aim of this paper is to outline a general strategy for studying top-quark
interactions in the context of an EFT, starting from the case of top-quark FCNC
processes. Our main points can be summarized as follows:
\begin{itemize}\itemsep0pt
	\item The widely used formalism that relies on dimension-four and -five
	      operators in the electroweak (EW) broken phase is inadequate in
	      several respects.
	\item Calculations of FCNC processes can now be performed (in most cases
	      already automatically) in the EFT framework at NLO in QCD. Some
	      new NLO results for four-fermion operator contributions are
	      provided here for the first time.
	\item A consistent analysis should be global, {\ie}, consider all
	      operators contributing to a given process. For such an approach to
	      be successful a sufficiently large (and complete) set of
	      observables should be identified. We show that for FCNC
	      interactions involving the top quark this is already close to being
	      possible with the current measurements and suggest a minimal set
	      of observables accessible at the LHC to complete the set.
\end{itemize}

The paper is organized as follows. \autoref{sec:mixing} discusses the operators
mixing effects at NLO in QCD and demonstrates the need for a global approach. We
show some NLO results for single-top production processes in \autoref{sec:nlo},
including both two- and four-fermion operators. A first global analysis
incorporating the most sensitive experimental searches is finally carried
out in \autoref{sec:global}.

\section{Effective field theory}\label{sec:eft}
Let us start, in this \autoref{sec:eft}, by presenting the effective operators
relevant for a NLO description of top-quark FCNC processes, and highlighting
the insufficiencies of the dimension-four and -five operators formalism.

\subsection{Fully gauge-invariant operators}

Assuming the full standard model $SU(3)_C\times SU(2)_L\times U(1)_Y$ gauge
symmetry as well as baryon and lepton number conservations,\footnote{See
Ref.~\cite{Dong:2011rh} for an EFT discussion of the baryon-number-violating
interactions of the top quark.} the first beyond-the-standard-model operators
$O_i$ constructed with standard-model fields only arise at dimension six.
Restricting our EFT description to this level, the
 Lagrangian can be written~\cite{Buchmuller:1985jz}:
\begin{equation}
  \lag{EFT}=\lag{SM}+\sum_i\frac{C_i}{\Lambda^2}O_i,
\label{eq:eff_lag}
\end{equation}
where $C_i$'s are dimensionless coefficients and $\Lambda$ is a mass scale. We will
use the operator basis and notations of Ref.~\cite{Grzadkowski:2010es}, which
includes 59 independent dimension-six operators. Our choice of operator
normalization follows Ref.~\cite{Zhang:2014rja}. 

Amongst the ones contributing (up to NLO in QCD) to top-quark FCNC processes,
different categories can be distinguished. We first consider operators involving
exactly two quarks. Their Lorentz structures can be used to separate three
subclasses: vector, scalar, and tensor operators. Omitting indices for clarity
(notably flavor ones) and denoting the fermionic flavor-generic gauge
eigenstates by $\ges q$, $\ges u$, $\ges d$, $\ges l$ and $\ges e$, they are:
\begin{align*}
& O_{\varphi q}^{1}\equiv \frac{y_t^2}{2}\;
	\bar{\ges q}\gamma^\mu \ges q	\;\;
	\FDF[i]	,
\quad O_{\varphi q}^{3}\equiv \frac{y_t^2}{2}\;
	\bar{\ges q}\gamma^\mu\tau^I\ges q	\;\;
	\FDFI[i]	,
\\[-1mm]
&O_{\varphi u}\equiv \frac{y_t^2}{2}\;\;
	\bar{\ges u}\gamma^\mu \ges u	\quad
	\FDF[i]	,
\\[2mm]
&O_{u\varphi}\equiv -y_t^3\quad
	\bar{\ges q}\ges u	\;
	\tilde\varphi	\quad
	(\varphi^\dagger\varphi -v^2/2)	,
\\[2mm]
&O_{uB}\equiv y_t g_Y\quad
	\bar{\ges q}\sigma^{\mu\nu}\ges u	\;
	\tilde{\varphi}	\quad
	B_{\mu\nu}	,
\\
&O_{uW} \!\equiv y_t g_W\;\;
	\bar{\ges q}\sigma^{\mu\nu}\tau^I\ges u	\,
	\tilde{\varphi}	\;\;
	W^I_{\mu\nu}	,
\\[-.5mm]
&O_{uG}\equiv y_t g_s\quad
	\bar{\ges q}\sigma^{\mu\nu}T^A\ges u	\;
	\tilde{\varphi}	\quad
	G^A_{\mu\nu}	,
\end{align*}
with $\overleftrightarrow D_\mu^{(I)}\equiv (\tau^I)\overrightarrow D_\mu -
\overleftarrow D_\mu(\tau^I)$ and $\tilde \varphi\equiv i\tau^2\varphi^*\equiv
\varepsilon\varphi^*$. Out of the $O_{\varphi
q}^{\pm}\equiv O_{\varphi q}^1\pm O_{\varphi q}^3$ operators, only $O_{\varphi
q}^-$ contributes to FCNC processes in the up sector. 
We note that the vector contributions to the $tqZ$ vertices arising from
$O_{\varphi q}^{-}$ and $O_{\varphi u}$, and the tensor
ones arising from $O_{uB}$ and $O_{uW}$ have not been both simultaneously
considered in experimental searches so far. Next we consider operators
involving two quarks and two leptons:
\begin{align*}
&
O^{1}_{lq}\equiv
	\bar{\ges l}\gamma_\mu \ges l	\quad
	\bar{\ges q}\gamma^\mu \ges q	,\qquad&&
O^{3}_{lq}\equiv
	\bar{\ges l}\gamma_\mu\tau^I \ges l	\quad
	\bar{\ges q}\gamma^\mu\tau^I \ges q	,\\[-1mm]&
O_{lu}\equiv
	\bar{\ges l}\gamma_\mu \ges l	\quad
	\bar{\ges u}\gamma^\mu \ges u	,\\&
O_{eq}\equiv
	\bar{\ges e}\gamma^\mu \ges e	\quad
	\bar{\ges q}\gamma_\mu \ges q	,\\&
O_{eu}\equiv
	\bar{\ges e}\gamma_\mu \ges e	\quad
	\bar{\ges u}\gamma^\mu \ges u	
,\\[2mm]
&
O_{lequ}^{1}\equiv
	\bar{\ges l}\ges e	\;\;\varepsilon\;\;
	\bar{\ges q}\ges u	,
\\[2mm]&
O_{lequ}^{3}\equiv
	\bar{\ges l}\sigma_{\mu\nu}\ges e	\quad\varepsilon\quad
	\bar{\ges q}\sigma^{\mu\nu}\ges u	.
\end{align*}
It is useful to introduce the $O_{lq}^\pm\equiv O_{lq}^1\pm O_{lq}^3$ combinations.
$O_{lq}^-$ contains the interactions of two up-type quarks with two charged leptons
(or, of two down-type quarks with two neutrinos) while $O_{lq}^+$ notably gives rise to
interactions between two up-type quarks and two neutrinos (or, two down-type
quarks and two charged leptons).
Finally, there are four-quark operators. The complete basis has been discussed in
Ref.~\cite{AguilarSaavedra:2010zi}. Here we use the basis of
Ref.~\cite{Grzadkowski:2010es} in which there are no tensor operators:
\begin{align*}
&
O^{1}_{qq}\equiv
	\bar{\ges q}\gamma_\mu \ges q	\quad
	\bar{\ges q}\gamma^\mu \ges q	,\qquad&&
O^{3}_{qq}\equiv
	\bar{\ges q}\gamma_\mu\tau^I \ges q	\quad
	\bar{\ges q}\gamma^\mu\tau^I \ges q	,\\&
O^{1}_{qu}\equiv
	\bar{\ges q}\gamma_\mu \ges q	\quad
	\bar{\ges u}\gamma^\mu \ges u	,\qquad&&
O^{8}_{qu}\equiv
	\bar{\ges q}\gamma_\mu T^A \ges q	\quad
	\bar{\ges u}\gamma^\mu T^A \ges u	,\\[-.5mm]&
O^{1}_{qd}\equiv
	\bar{\ges q}\gamma_\mu \ges q	\quad
	\bar{\ges d}\gamma^\mu \ges d	,\qquad&&
O^{8}_{qd}\equiv
	\bar{\ges q}\gamma_\mu T^A \ges q	\quad
	\bar{\ges d}\gamma^\mu T^A \ges d	,\\&
O_{uu}\equiv
	\bar{\ges u}\gamma_\mu \ges u	\quad
	\bar{\ges u}\gamma^\mu \ges u	,\\[-.5mm]&
O^{1}_{ud}\equiv
	\bar{\ges u}\gamma_\mu \ges u	\quad
	\bar{\ges d}\gamma^\mu \ges d	,\qquad&&
O^{8}_{ud}\equiv
	\bar{\ges u}\gamma_\mu T^A \ges u	\quad
	\bar{\ges d}\gamma^\mu T^A \ges d
,\\[2mm]\displaybreak[0]
&
O^{1}_{quqd}\equiv
	\bar{\ges q}\ges u	\;\;\varepsilon\;\;
	\bar{\ges q}\ges d	,\qquad&&
O^{8}_{quqd}\equiv
	\bar{\ges q}\, T^A \ges u	\quad\varepsilon\quad
	\bar{\ges q}\, T^A \ges d	.
\end{align*}
The Hermitian conjugates of scalar and tensor operators need to be
added to those three lists of two- and four-fermion operators while the imposed
Hermiticity, $C\ind ab = \big( C\ind ba\big)^*$, of vector operator coefficients
ensures the reality of the effective Lagrangian.

The gauge invariance of the SM could only be imposed on fermionic gauge
eigenstates. Observations, however, are related to mass eigenstates. Rotating
one basis to the other is therefore required for all practical purposes. Neither
the gauge-eigenstate operator coefficients, nor the unitary
rotation matrices appearing in Yukawa singular-value decompositions are measurable.
Physical operator coefficients and CKM as well as PMNS mixing matrix
elements are.
\newcommand*{\phiq}{C^{1}_{\varphi q}}%
\newcommand*{\phiqt}{\tilde C^{1}_{\varphi q}}%
\newcommand*{\phiqp}{C_{\varphi q}^{1}\!\!'}%
\newcommand{\ckm}{V_\text{\tiny CKM}}%
\newcommand{\pmns}{V_\text{\tiny PMNS}}%
By removing unphysical rotation matrices, we choose the
gauge-eigenstates to be expressed in terms of physical eigenstates as
\begin{gather*}
\ges q\equiv (u_L,\; \ckm d_L)^T, \qquad
\ges u\equiv u_R,	\qquad
\ges d\equiv d_R,\\
\ges l \equiv (\pmns \nu_L,\; e_L)^T,	\qquad
\ges e \equiv e_R.
\end{gather*}
The physical operator coefficients involving left-handed up- and down-type quarks are then related through the CKM matrix. For instance, in the $O^1_{\varphi q}$ case,
\begin{equation}
\begin{aligned}
\bar{\ges q}\gamma^\mu \phiq\ges q 
&=	\bar u_L\gamma^\mu\; \phiq\; u_L\\
&+	\bar d_L\gamma^\mu\; [\ckm^\dagger \phiq \ckm]\; d_L.
\end{aligned}
\label{eq:ud_correlation}
\end{equation}
Similarly, the coefficients of operators involving left-handed
charged leptons and neutrinos are related to each other through the PMNS matrix.

Putting quark flavor indices between brackets, there are ten independent
complex coefficients, for either $a=1$ or $2$, in the two-quark category:
\begin{gather*}
C_{\varphi q}^{-(a3)} = C_{\varphi q}^{-(3a)*} \;\equiv C_{\varphi q}^{-(a+3)}, \\
C_{\varphi u}^{ (a3)} =	C_{\varphi u}^{ (3a)*} \;\equiv C_{\varphi u}^{ (a+3)}, \\
C_{u\varphi}^{(a3)}	,\; C_{u\varphi}^{(3a)}, \\
C_{uB}^{(a3)}		,\; C_{uB}^{(3a)}, \qquad
C_{uW}^{(a3)}		,\; C_{uW}^{(3a)}, \qquad
C_{uG}^{(a3)}		,\; C_{uG}^{(3a)}.
\end{gather*}
Without distinguishing the lepton flavors (all diagonal and nondiagonal
combinations should in principle be considered independently), there are nine
operators, for
either $a=1$ or $2$, in the two-quark--two-lepton category:
\begin{gather*}
C_{lq}^{-(a3)} = C_{lq}^{-(3a)*} \;\equiv C_{lq}^{-(a+3)}, \\
C_{lq}^{+(a3)} = C_{lq}^{+(3a)*} \;\equiv C_{lq}^{+(a+3)}, \\
C_{lu}^{(a3)} = C_{lu}^{(3a)*} \;\equiv C_{lu}^{(a+3)}, \\
C_{eq}^{(a3)} = C_{eq}^{(3a)*} \;\equiv C_{eq}^{(a+3)}, \\
C_{eu}^{(a3)} = C_{eu}^{(3a)*} \;\equiv C_{eu}^{(a+3)}, \\
C_{lequ}^{1(a3)}	,\; C_{lequ}^{1(3a)}, \qquad
C_{lequ}^{3(a3)}	,\; C_{lequ}^{3(3a)}.
\end{gather*}
Finally, for each allowed combination of $a$, $b$, $c\in\{1,2\}$, there are
$11$ independent four-quark coefficients leading to top FCNC processes:
\begin{gather*}
C_{qq}^{1(3a,bc)} = C_{qq}^{1(bc,3a)} = C_{qq}^{1(a3,cb)*}, \\
C_{qq}^{3(3a,bc)} = C_{qq}^{3(bc,3a)} = C_{qq}^{3(a3,cb)*}, \\
C_{uu}^{(3a,bc)} = C_{uu}^{(bc,3a)} = C_{uu}^{(a3,cb)*}, \\
C_{ud}^{1(3a,bc)} = C_{ud}^{1(a3,cb)*}, \qquad
C_{ud}^{8(3a,bc)} = C_{ud}^{8(a3,cb)*}, \\
C_{qu}^{1(3a,bc)} = C_{qu}^{1(a3,cb)*}, \qquad
C_{qu}^{8(3a,bc)} = C_{qu}^{8(a3,cb)*}, \\
C_{qu}^{1(bc,3a)} = C_{qu}^{1(cb,a3)*}, \qquad
C_{qu}^{8(bc,3a)} = C_{qu}^{8(cb,a3)*}, \\
C_{qd}^{1(3a,bc)} = C_{qd}^{1(a3,cb)*}, \qquad
C_{qd}^{8(3a,bc)} = C_{qd}^{8(a3,cb)*}.
\end{gather*}
Four-fermion operators have been overlooked in most analyses.

\subsection{Dimension-four and -five operators}

Beside the fully gauge-invariant effective field theory that will be exploited
here, different theoretical frameworks have been used in the literature to
describe top-quark flavor-changing neutral currents in a model-independent way.

A very common approach is the anomalous coupling one. Its main advantage is that
of being close to the Feynman rules definition and so, of easy use.
It is, however, not a well-defined quantum field theory where constraints set by
symmetries and radiative corrections can be taken into account systematically.
Such an approach is therefore not suitable for the purpose of a global analysis
at next-to-leading order in QCD.

Second, an effective-field-theory description of top-quark FCNCs in the
electroweak broken phase~\cite{Beneke:2000hk, AguilarSaavedra:2004wm} has been
widely used. It is based on an effective Lagrangian containing
dimension-four and -five operators that only satisfy Lorentz and
$SU(3)_C\times U(1)_{EM}$ gauge symmetries. This broken-phase effective
Lagrangian reads
\begin{equation}
\begin{aligned}
\lag{eff}^{\cancel{EW}}=
&
-\frac{g_W}{2c_W}\quad
	\bar{t}\gamma^\mu(v_{tq}^Z-a_{tq}^Z\gamma_5)q\quad
	Z_\mu\\&
-\frac{g_W}{2\sqrt{2}}g_{qt}\quad
	\bar{q}(g_{qt}^v+g_{qt}^a\gamma_5)t\quad
	h\\&
-e\frac{\kappa^\gamma_{tq}}{\Lambda}\quad
	\bar{t}\sigma^{\mu\nu} (f_{tq}^\gamma+ih_{tq}^\gamma\gamma_5)q\quad
	A_{\mu\nu}\\&
-\frac{g_W}{2c_W}\frac{\kappa_{tq}^Z}{\Lambda}\quad
	\bar{t}\sigma^{\mu\nu} (f_{tq}^Z+ih_{tq}^Z\gamma_5)q\quad
	Z_{\mu\nu}\\&
-g_s\frac{\kappa_{tq}^g}{\Lambda}\quad
	\bar{t}\sigma^{\mu\nu}T^A (f_{tq}^g+ih_{tq}^g\gamma_5)q\quad
	G_{\mu\nu}^A\\&
+\hc
\end{aligned}
\label{eq:eff45}
\end{equation}
where $q=u$ or $c$ and $c_W\equiv\cos\theta_W$. The $\kappa_{tq}$ and $g_{qt}$
coefficients are real and positive, while the complex $(v^Z_{tq},a^Z_{tq})$,
$(f_{tq},h_{tq})$ and $(g^v_{qt},g^a_{qt})$ pairs satisfy:
$|f_{tq}|^2+|h_{tq}|^2=1$ and $|g^v_{tq}|^2+|g^a_{tq}|^2=1$.

Without further constraint on its parameters, such an effective Lagrangian may
be understood as more general than the fully gauge-invariant one. In other
words, new physics is not assumed to preserve the electroweak
$SU(2)_L\times U(1)_Y$ gauge symmetry. Such a construction turns out to be 
way too general as it would require many coefficients to be extremely small or
correlated with each other in order to reproduce measurements.
This is at variance with the full $SU(3)_C\times SU(2)_L\times U(1)_Y$ symmetry
that naturally accounts for all observations made so far. For example, with full
gauge invariance imposed, flavor-changing neutral currents only occur at the 
loop or nonrenormalizable level.

In this work, we assume that new physics preserves the full
standard-model gauge invariance, at least approximately, in the energy range
probed by the LHC. The Lagrangian of \autoref{eq:eff45} may then be seen as a
practical reparametrization of the $SU(3)_C\times SU(2)_L\times U(1)_Y$-invariant
operators presented in the previous section. Its couplings are then tacitly
understood as expressions of the fully gauge-invariant operator coefficients,
the SM parameters and the scale $\Lambda$:
\begin{align*}
-\frac{g_W}{2c_W} 
\Big\{\begin{array}{@{}l@{}}	v^Z_{tq}	\\ -a^Z_{tq}	\end{array}
	&= \frac{-e}{2s_Wc_W} \frac{m_t^2}{\Lambda^2}
	\big[
	C_{\varphi u}^{(a+3)*}
	\pm C_{\varphi q}^{-(a+3)*}
	\big],
\\
-\frac{g_W}{2\sqrt{2}} g_{qt} 
\Big\{\begin{array}{@{}l@{}}	g_{qt}^v	\\  g_{qt}^a	\end{array}
	&= \frac{-2m_t}{v}\frac{m_t^2}{\Lambda^2}
	\left[
	C_{u\varphi}^{(a3)}
	\pm C_{u\varphi}^{(3a)*}
	\right],
\\
-e\frac{\kappa^\gamma_{tq}}{\Lambda}
\Big\{\begin{array}{@{}l@{}}	f^\gamma_{tq}	\\[1mm] ih^\gamma_{tq}	\end{array}
	&= e \frac{m_t}{\Lambda^2}
	\begin{aligned}[t]
	\big[
	  &(C_{uB}^{(3a)}+C_{uW}^{(3a)})\\
	\pm&(C_{uB}^{(a3)}+C_{uW}^{(a3)})^*
	\big],
	\end{aligned}
\\
-\frac{g_W}{2c_W} \frac{\kappa^Z_{tq}}{\Lambda}
\Big\{\begin{array}{@{}l@{}}	f^Z_{tq}	\\[1mm]	 ih^Z_{tq}	\end{array}
	&= \frac{-e}{s_Wc_W} \frac{m_t}{\Lambda^2}
	\begin{aligned}[t]
	\big[
	 &(s_W^2\; C_{uB}^{(3a)} -c_W^2\; C_{uW}^{(3a)})\\
	\pm&(s_W^2\; C_{uB}^{(a3)} -c_W^2\; C_{uW}^{(a3)})^*
	\big],
	\end{aligned}
\\
-g_s \frac{\kappa^g_{tq}}{\Lambda}
\Big\{\begin{array}{@{}l@{}}	f^g_{tq} 	\\[1mm] i h^g_{tq}	\end{array}
	&= {g_s}\frac{m_t}{\Lambda}
	\big[
	  C_{uG}^{(3a)}
	\pm C_{uG}^{(a3)*}
	\big].
\end{align*}%
Such a reparametrization has, however, intrinsic limitations and to some extent
can lead to misconceptions.  Let us list a few specific reasons.

First, the broken-phase effective Lagrangian displayed as in \autoref{eq:eff45}
hides the actual scaling of each contribution. It does not make explicit that
FCNC operators do not appear at the renormalizable level, and therefore it does
not account for the experimental lack of evidence of the corresponding effects
in the first place. The hierarchies this Lagrangian displays are moreover
misleading as all tree-level FCNC effects actually first appear at dimension
six. The operators that are seemingly of dimension four and five actually
contribute at the same order in $1/\Lambda$. At next-to-leading order in QCD,
the $\bar t\sigma^{\mu\nu}T^Aq\; G_{\mu\nu}^A$ operators renormalize the $\bar t
q h$ ones that, in the broken phase, seem to be of different dimension. 
On the contrary, in the fully gauge-invariant picture, $O_{uG} \equiv \bar{\ges q}
\sigma^{\mu\nu}T^A\ges u\; \tilde\varphi\; G_{\mu\nu}^A$ renormalizes
$O_{u\varphi} \equiv \bar{\ges q} \ges u\; \tilde\varphi\;
\varphi^\dagger\!\varphi$ without actually mixing dimensions. More details on
the mixing between operators and their running will be provided in the next
section. Note in passing that the use of a broken-phase effective Lagrangian
would make the computation of NLO weak corrections intractable.

\begin{figure}[tb]\centering
	\begin{subfigure}{\columnwidth}
		\centering
		\includegraphics[width=6cm]{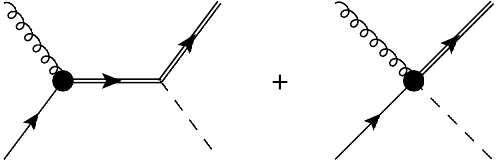}
		\caption{}
		\label{fig:ugth}
	\end{subfigure}
	\\[3mm]
	\begin{subfigure}{\columnwidth}
		\fmfframe(2,2)(2,2){
		\begin{fmfgraph*}(30,16)
			\fmfleft{l1,l2}
			\fmfright{r1,r2}
			\fmf{fermion}{l1,v,l2} 
			\fmf{photon, tens=1.0}{v,w}
			\fmf{fermion}{r1,w}
			\fmf{top}{w,r2}
			\fmfv{d.shape=square,d.size=6pt,d.filled=30}{w}
			\fmffreeze
			\fmfcmd{
				label.bot(btex $\gamma, Z$ etex, .5[vloc(__v),vloc(__w)]);
				label.lft(btex $e^-$ etex, vloc(__l1));
				label.lft(btex $e^+$ etex, vloc(__l2));
				label.rt(btex $t$ etex, vloc(__r2));
				label.rt(btex $bar u$ etex, vloc(__r1));
			}
		\end{fmfgraph*}
		}
		\raisebox{8.5mm}{ + }
		\fmfframe(4,2)(2,2){
		\begin{fmfgraph*}(20,16)
			\fmfleft{l1,l2}
			\fmfright{r1,r2}
			\fmf{fermion}{l1,v,l2}
			\fmf{fermion}{r1,v}
			\fmf{top}{v,r2}
			\fmfv{d.shape=triangle,d.size=10pt,d.filled=30}{v}
			\fmffreeze
			\fmfcmd{
				label.lft(btex $e^-$ etex, vloc(__l1));
				label.lft(btex $e^+$ etex, vloc(__l2));
				label.rt(btex $t$ etex, vloc(__r2));
				label.rt(btex $bar u$ etex, vloc(__r1));
			}
		\end{fmfgraph*}
		}
		\caption{}
		\label{fig:eetu}
	\end{subfigure}
	\caption{The full standard-model gauge symmetry gives rise to four-point
	interactions, not included in the Lagrangian of \autoref{eq:eff45}, that
	contribute at the same order as three-point ones in some FCNC processes:
	\emph{e.g.} in $ug\to th$ production (or radiative $t\to hug$ decay), or
	in $e^+e^-\to t\,\bar u$ (and $t\to u\,e^+e^-$).}
	\label{fig:missing_four_point_operators}
\end{figure}

Second, some operators contributing at the same order as those appearing in
\autoref{eq:eff45} are not included. Namely, a $\bar t\sigma^{\mu\nu}T^Aq \; h\;
G_{\mu\nu}^A$ operator actually contributes to $th$ production at hadron
colliders at the same order as $\bar t\sigma^{\mu\nu}T^Aq\; G_{\mu\nu}^A$ (see
\autoref{fig:ugth}). This is trivially seen when the full gauge invariance is
restored as both contributions then arise from the same dimension-six
$O_{uG}$ operator.

A complete basis should also include four-fermion operators. These are also of
dimension six and can be related to two-fermion operators through the equations
of motion (EOM). Such \emph{contact} interactions could arise, for instance, in
the presence of a heavy mediator coupling to two fermionic currents. They have
unduly been neglected in experimental searches, Refs.~\cite{Achard:2002vv, DELPHI:2011ab}
excepted. They could for example contribute to processes such as 
$t\to j\,\ell^+\ell^-$, $pp\to tj$, $e^+e^-\to t j$ (\autoref{fig:eetu}) and
$e^- p \to e^-t$. Trying to use the equations of motion to trade them all for
two-fermion operators involving more covariant derivatives is in fact vain.
Those involving vector or tensor fermionic bilinears which are not flavor
diagonal would not appear in any of the EOM (non-flavor-diagonal scalar
bilinears, on the other hand, are present in the equations of motion for the
Higgs field).
 In many interesting leading-order processes and in all
next-to-leading-order ones, the off-shell character of the particles involved in an
effective operator also precludes the use of EOM that could render
irrelevant some operators containing derivatives.
In \autoref{fig:missing_four_point_operators} some examples of
processes proceeding through the exchange of off-shell particles are provided.
All dominant effects of heavy new physics can therefore only be guaranteed 
to be modeled by an effective theory if four-fermion operators are included.

Third, by writing an effective theory in the electroweak broken phase without
making explicit the expressions of the couplings in terms of the fully
gauge-invariant operator coefficients, one can easily overlook correlations
between operators. The $\bar t\sigma^{\mu\nu}T^Aq \; h\; G_{\mu\nu}^A$ and $\bar
t\sigma^{\mu\nu}T^Aq\; G_{\mu\nu}^A$ interactions that derive from the same
dimension-six $O_{uG}$ operator already provided an obvious example. Such kind
of full correlation due to the presence, in $\varphi$, of a physical Higgs
particle and a vacuum expectation value occurs only above the electroweak
symmetry breaking scale in processes involving an external Higgs particle or
when taking loop-level contributions into account. Another type of correlation
arises from the fact that left-handed down- and up-type quarks belong to a single
gauge-eigenstate doublet. Operator coefficients measurable in $B$-meson physics
(see Refs.~\cite{Alonso:2014csa, Buras:2014fpa} for recent EFT analyses
exploiting the full standard-model gauge invariance) are actually related to
those relevant to top-quark physics through equalities like
\autoref{eq:ud_correlation} that involve $\ckm$. The impact of $B$-physics
constraints on top FCNC operators has for instance been studied in
Refs.~\cite{Fox:2007in, Li:2011fza, Li:2011af, Gong:2013sh} with one single
operator switched on at the time, though. A truly global analysis in a fully
gauge-invariant EFT framework remains to be carried out. It should take advantage
of all types of correlations and use several processes in which a closed set of
operators contributes through different combinations. Only by doing this, is it
possible to disentangle the effects coming from each of them.

\section{Mixings}\label{sec:mixing}

At NLO in QCD, dimension-six operators may mix with each other. In other words,
the renormalization of an operator may involve several others. Because of quantum
effects, the very definition of operator coefficients actually depends on the
renormalization scheme and scale. RG mixings imply that a coefficient
defined at one scale is actually a combination of many others at a
different scale. The information about the RG flow,
\begin{flalign}
  \frac{\D C_i(\mu)}{\D \ln\mu}=\gamma_{ij}C_j(\mu),
\end{flalign}
is encoded in the anomalous dimension matrix $\gamma_{ij}$ which has been
computed recently for the full set of dimension-six
operators~\cite{Jenkins:2013zja, Jenkins:2013wua, Alonso:2013hga,
Alonso:2014zka} (see also Ref.~\cite{Zhang:2014rja} for a specific discussion of
the anomalous dimensions of flavor-changing top-quark operators).

At a high scale, a full theory may justify specific values for operators
coefficients. In particular, some of them may be negligible. However, after the
RG evolution down to lower scales, mixing might lead to a significant increase of the set
of operators with sizable coefficients.
As an example, let us consider the Yukawa operator $O_{u\varphi}^{(13)}$. It can be
be generated from its QCD mixing only by the color-dipole operator
$O_{uG}^{(13)}$.  Over a range of scales as small as $1$~TeV$\to m_t$, one gets
\begin{equation*}\left\{
\begin{aligned}
	C_{uG}^{(13)}(1~\text{TeV})=1, \\
	C_{u\varphi}^{(13)}(1~\text{TeV})=0,
\end{aligned}\right.
\quad\longrightarrow\quad\left\{
\begin{aligned}
	C_{uG}^{(13)}(m_t)=0.98, \\
	C_{u\varphi}^{(13)}(m_t)=0.23 .
\end{aligned}\right.
\end{equation*}
At the energies currently probed by experiments, it is thus
unnatural to assume only one or two operator coefficients are nonzero. One
should \emph{a priori} include all operators contributing at a given order. Then,
to constrain operator coefficients consistently, one should use the
renormalization-group equations and evolve all available bounds to a common
scale where a global analysis can then be carried out.

\begin{figure*}[tb]\centering
\includegraphics[width=.9\linewidth]{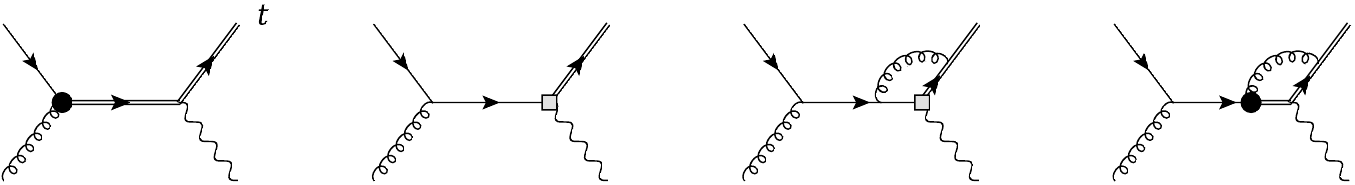}
\caption{Representative $ug\to tZ$ diagrams involving $O_{uG}$ (black dot) and
$O_{uW}$ (gray square) operators, at leading order (first two) and
next-to-leading order (last two diagrams). The UV divergence of the fourth
diagram involving $O_{uG}$ is regularized by a counterterm of $O_{uW}$ form.}
\label{fig:mixorder}
\end{figure*}

\subsection{Renormalization patterns}

The mutual renormalizations of operator coefficients entail that, at
next-to-leading order (and beyond), some operators will provide counterterms
regularizing UV divergent diagrams involving other ones.
As an example, let us consider the $ug\to tZ$ production process. Some
representative diagrams are given in \autoref{fig:mixorder}. The first two
are leading-order amplitudes involving $O_{uG}$ and $O_{uW}$ operators,
respectively. The third and fourth diagrams provide $\mathcal{O}(\alpha_s)$
corrections to the second and first. The fourth contribution also requires
a counterterm from $O_{uW}$. On the contrary, at NLO in QCD, there is no
divergent diagram involving $O_{uW}$ that would require a counterterm of
$O_{uG}$ form.

The pattern of such mutual NLO renormalizations
can in principle be extracted from the RG equations of
Refs.~\cite{Jenkins:2013zja, Jenkins:2013wua, Alonso:2013hga, Alonso:2014zka}.
The full anomalous dimension matrix is, however, complicated and obtaining this
information may appear nontrivial. Changing the normalization of operator
coefficients so as to make their LO contributions formally of the same order
renders the situation clearer. Some pieces of the RG equations for the new
coefficients then formally appear of order $\alpha_s$ and can thus be isolated. They
contain the information about renormalization patterns we need. 

Having taken the normalization (with $y_t$ and $g_{Y,W,s}$ factors) of our
two-quark operators as in Ref.~\cite{Zhang:2014rja},
and assuming for the moment all coefficients to have comparable magnitudes makes
their LO contributions to the $pp\to t\gamma$, $tZ$ and $th$ processes formally
of the same order. Such a normalization appears natural if one considers that
each boson is eventually attached to a fermionic line in the full theory from
which the EFT models the low-energy effects. One then obtains a closed set of RG
equations for $C_{u\varphi}$, $C_{uB}$, $C_{uW}$ and $C_{uG}$ that are formally
of order $\alpha_s$. The corresponding anomalous dimension matrix is given by
\newcommand{\bla}{}%
\begin{equation*}
\renewcommand{\arraystretch}{1.3}%
  \frac{2\alpha_s}{\pi}\left( 
  \begin{array}{ccccccc}
    -2	& 0	& 0	& -1	\\
    0	& 1/3	& 0	& 5/9	\\
    0	& 0	& 1/3	& 1/3	\\
    0	& 0	& 0	& 1/6
  \end{array}
  \right).
\end{equation*}
Note that, due to current conservation, the coefficients of the vector $O_{\varphi
q}^{-}$ and $O_{\varphi u}$ operators are not renormalized.

It is then transparent that $O_{uG}$ renormalizes all other coefficients which,
on the contrary, only renormalize themselves at order $\alpha_s$. For instance,
the RG equation of $\bla C_{uW}^{(13)}$ at that order reads
\begin{align*}
  \frac{\D \bla C_{uW}^{(13)}(\mu)}{\D \ln\mu}=
  \frac{2\alpha_s}{3\pi} \bla C_{uW}^{(13)}(\mu)+ 
  \frac{2\alpha_s}{3\pi} \bla C_{uG}^{(13)}(\mu).
\end{align*}
The first term corresponds to the running of $\bla C_{uW}^{(13)}$ itself, while
the second one is a mixing from $\bla C_{uG}^{(13)}$ to $\bla C_{uW}^{(13)}$
formally of order $\alpha_s$ which will cancel the UV divergences from
the fourth diagram (and of diagrams not shown in \autoref{fig:mixorder}). On
the other hand, $\bla C_{uG}^{(13)}$ is not renormalized by $\bla
C_{uW}^{(13)}$:
\vspace*{-2mm}
\begin{align*}
  \frac{\D \bla C_{uG}^{(13)}(\mu)}{\D \ln\mu}
  =\frac{\alpha_s}{3\pi} \bla C_{uG}^{(13)}(\mu)
\end{align*}
at order $\alpha_s$.
 
\begin{figure}[tb]\centering
\includegraphics[height=2cm]{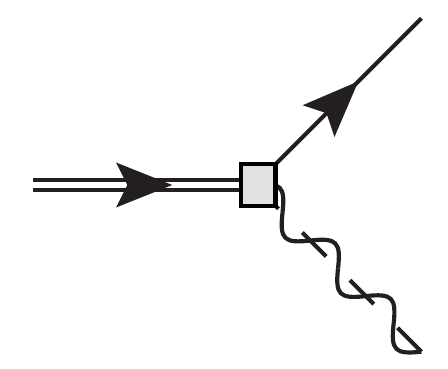}\qquad
\includegraphics[height=2cm]{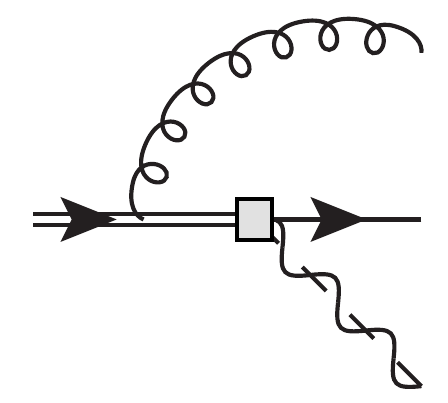}\qquad
\includegraphics[height=2cm]{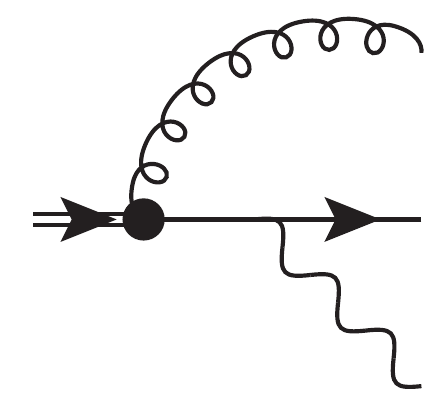}
\caption{Representative diagrams for $t\to u\gamma$, $qZ$, $qh$ at leading and
next-to-leading orders. The gray square represents a contribution from the
$O_{uW}$ weak-dipole operator, while the black dot represents a contribution
from the $O_{uG}$ color-dipole operator.}
\label{fig:ugmixing}
\end{figure}

At leading order, there is only one contribution to the $t\to q\gamma$, $qZ$ and
$qh$ decay processes: the first diagram of \autoref{fig:ugmixing}. The above
procedure therefore does not apply. The second diagram is a QCD correction to the
first one. Other quantum corrections involving $O_{uG}$, like the third diagram,
are also included by analogy.

As four-fermion operators could arise from the tree-level exchange of a new
heavy bosonic mediator, their coefficients could probably be chosen of
new-physics order only. Examining the RG equations at order $\alpha_s$, it turns
out that, unlike four-quark operators, two-quark--two-lepton ones do not mix
between themselves (this may not be true in another basis) or with two-quark
operators. The pattern of renormalization of two-quark and two-quark--two-lepton
operators by four-quark ones is detailed in \autoref{tab:mixings_four_quarks}.

\begin{table}[tb]
\centering
\begin{tabular*}{.9\columnwidth}{@{\extracolsep{\fill}}cccccc}
\hline\noalign{\vskip 1mm}
	& $O_{qq}^{1}$,$O_{qq}^{3}$ & $O_{qu}^{1}$ & $O_{ud}^{1}$
	& $O_{uu}$ & $O_{qd}^{1}$
	\\[1mm]\hline\noalign{\vskip 1mm}
	$O_{\varphi q}^{1}$ & $\checkmark$ & $\checkmark$ &&&\checkmark
	\\
	$O_{\varphi q}^{3}$ & $\checkmark$ &&&&
	\\
	$O_{\varphi u}$ & &$\checkmark$&$\checkmark$&$\checkmark$&
	\\[1mm]\hline\noalign{\vskip 1mm}
	$O_{lq}^{1}$ &$\checkmark$&$\checkmark$&&&$\checkmark$
	\\
	$O_{lq}^{3}$ &$\checkmark$&&&&
	\\
	$O_{lu}$&&$\checkmark$&$\checkmark$&$\checkmark$&
	\\
	$O_{eq}$&$\checkmark$&$\checkmark$&&&$\checkmark$
	\\
	$O_{eu}$&&$\checkmark$&$\checkmark$&$\checkmark$&
	\\[1mm]\hline
\end{tabular*}
\caption{Two-quark and two-quark--two-lepton operators that are renormalized by four-quark ones, at order $\alpha_s$.}
\label{tab:mixings_four_quarks}
\end{table}

\section{FCNC processes at NLO in QCD}\label{sec:nlo}

In order to demonstrate the feasibility of an EFT treatment of top-quark FCNC
processes that is NLO accurate in QCD, we discuss two specific cases. 
First, we consider top-quark decay processes and the ---often overlooked---
contributions of two-quark--two-lepton operators. Second, some NLO results
merged with parton shower are presented for single top produced in association
with a photon, $Z$ or Higgs bosons.

For numerical results, here and in what follows, we use: $m_t=172.5$\,GeV,
$\alpha^{-1}=127.9$, $\alpha_s=0.10767$, $\sin^2\theta_W=0.2337$,
$m_Z=91.1876$\,GeV and $\Gamma_Z=2.4952$\,GeV. Unless otherwise specified, we
set $\Lambda=1$\,TeV.

Note dimension-six effective operators not listed here contribute, at order
$1/\Lambda^2$, to the observables used to fix those SM parameters. As the
leading contributions to the FCNC processes we are considering appear at order
$1/\Lambda^4$, these shifts in the SM parameters only induce $1/\Lambda^6$
corrections which can be neglected consistently.

\subsection{Two-quark--two-lepton operators in top decay}

\begin{figure}[tb]
\includegraphics[width=.75\linewidth]{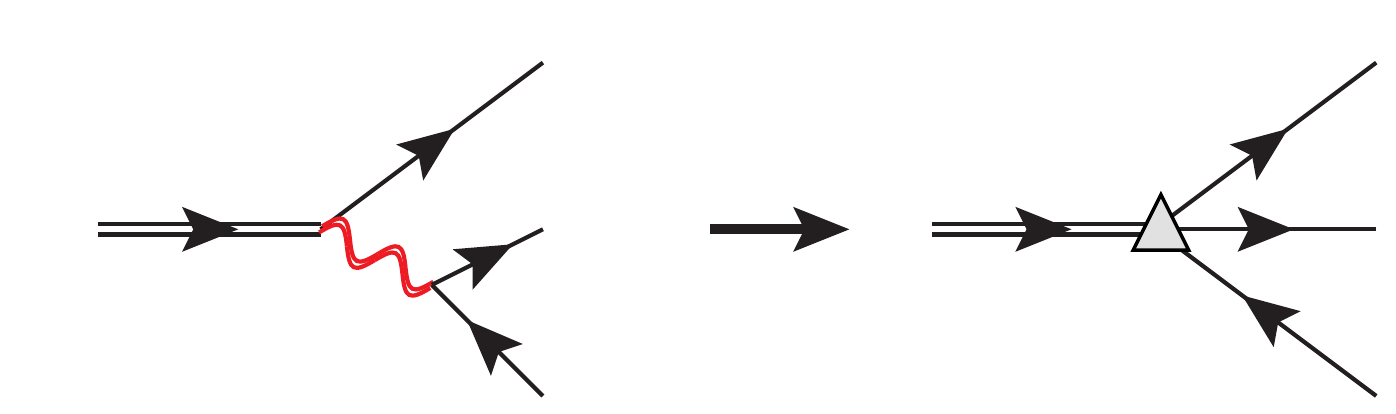}
\caption{The tree-level exchange of a heavy flavor-changing $Z'$ would generate
four-fermion effective operators. Such operators have been overlooked in many  experimental searches.}
\label{fig:shrink_4f}
\end{figure}

As stressed before, two-quark--two-lepton operators have been overlooked in top-quark
FCNC searches at the LHC even though they contribute to  $pp\to
t\,\ell^+\ell^-$, $t\to j\,\ell^+\ell^-$ or $e^+e^-\to tj$ processes (see
\autoref{fig:eetu}). As they are potentially tree-level generated, for instance
from the exchange of a heavy $Z'$ (see \autoref{fig:shrink_4f}), those operators
may even have coefficients larger than those of two-quark operators.

Complete results at NLO in QCD for top-quark FCNC decays through two-quark and
two-quark--two-lepton operators can be found in Ref.~\cite{Zhang:2014rja}. Here,
we focus on the specific $t\to j\,\ell^+\ell^-$ process. The two-quark operators
contribute through the exchange of a (virtual) photon or $Z$ while the
two-quark--two-lepton operators lead to three-body decays. The first half
of \autoref{tab:t_to_uee} presents the numerical contributions of each operator
coefficient to the partial decay width. As light-fermion masses have been neglected,
operators involving light fermions of different chiralities do not
interfere. Only the contributions of $O_{uG}$ (where a photon or $Z$ is
emitted from the quark line) differ between the left- and right-handed light quark
cases.

\begin{table*}[tbp]
\begin{align*}
\Gamma&_{t\to j\ell^+\ell^-}^\text{on-peak + off-peak}/10^{-5}\,\text{GeV}
\;\times (\Lambda/1\,\text{TeV})^4 =
\\
&\Re
\left(\begin{array}{@{}l@{}}
	C_{lq}^{-(a+3)}  \\[\n]
	C_{eq}^{(a+3)} \\[\n]
	C_{\varphi q}^{-(a+3)} \\[\n]
	C_{uB}^{(a3)}	\\[\n]
	C_{uW}^{(a3)} \\[\n]
	C_{uG}^{(a3)}
\end{array}\right)^{\hspace{-1.5mm}\dagger}
\left(
\begin{array}{*{5}{@{}c@{\hspace{3mm}}}c@{}}
 +\us{0.29}{-8}
	& 0
	& -\us{0.035}{-12} -\us{0.23}{-8}i
	& -\us{0.19}{-7} -\us{0.11}{-8}i
	& -\us{0.33}{-7} +\us{0.38}{-8}i
	& +\usv{0.026}{---} -\usv{0.0025}{---}i
\\
	& +\us{0.29}{-8}
	& +\us{0.028}{-12} +\us{0.18}{-8}i
	& -\us{0.25}{-7} +\us{0.087}{-8}i
	& -\us{0.14}{-7} -\us{0.30}{-8}i
	& +\usv{0.00064}{---} +\usv{0.023}{---}i
\\
	&
	& \us{+1.9}{-8}
	& +\us{1.8}{-8} -\us{0.016}{-8}i
	& -\us{6.2}{-8} -\us{0.016}{-8}i
	& +\usv{0.29}{---} +\usv{0.22}{---}i
\\
	&
	&
	& +\us{0.91}{-9}
	& -\us{3.6}{-9} -\us{0.049}{-9}i
	& +\usv{0.14}{---} +\usv{0.12}{---}i
\\
	&
	&
	&
	& +\us{7.6}{-9}
	& -\usv{0.61}{---} -\usv{0.55}{---}i
\\
	&
	&
	&
	&
	& +\usv{0.0068}{---}
\end{array}\right)
\left(\begin{array}{@{}l@{}}
	C_{lq}^{-(a+3)}  \\[\n]
	C_{eq}^{(a+3)} \\[\n]
	C_{\varphi q}^{-(a+3)} \\[\n]
	C_{uB}^{(a3)}	\\[\n]
	C_{uW}^{(a3)} \\[\n]
	C_{uG}^{(a3)}
\end{array}\right)
\\
+& \Re	
\left(\begin{array}{@{}l@{}}
	C_{lu}^{(a+3)}  \\[\n]
	C_{eu}^{(a+3)} \\[\n]
	C_{\varphi u}^{(a+3)} \\[\n]
	C_{uB}^{(3a)*}	\\[\n]
	C_{uW}^{(3a)*} \\[\n]
	C_{uG}^{(3a)*}
\end{array}\right)^{\hspace{-1.5mm}\dagger}
\left(
\begin{array}{*{5}{@{}c@{\hspace{3mm}}}c@{}}
+\us{0.29}{-8}
	& 0
	& -\us{0.035}{-12} -\us{0.23}{-8}i
	& -\us{0.19}{-7} -\us{0.11}{-8}i
	& -\us{0.33}{-7} +\us{0.38}{-8}i
	& +\usv{0.0068}{---} +\usv{0.021}{---}i
\\
	& +\us{0.29}{-8}
	& +\us{0.028}{-12} +\us{0.18}{-8}i
	& -\us{0.25}{-7} +\us{0.087}{-8}i
	& -\us{0.14}{-7} -\us{0.30}{-8}i
	& +\usv{0.016}{---} +\usv{0.0043}{---}i
\\
	& 
	& +\us{1.9}{-8}
	& +\us{1.8}{-8} -\us{0.016}{-8}i
	& -\us{6.2}{-8} -\us{0.016}{-8}i
	& -\usv{0.18}{---} -\usv{0.092}{---}i
\\
	& 
	& 
	& +\us{0.91}{-9}
	& -\us{3.6}{-9} -\us{0.049}{-9}i
	& -\usv{0.13}{---} -\usv{0.096}{---}i
\\
	& 
	& 
	& 
	& +\us{7.6}{-9}
	& +\usv{0.31}{---} +\usv{0.19}{---}i
\\
	& 
	& 
	& 
	& 
	& +\usv{0.0053}{---}
\end{array}\right)
\left(\begin{array}{@{}l@{}}
	C_{lu}^{(a+3)}  \\[\n]
	C_{eu}^{(a+3)} \\[\n]
	C_{\varphi u}^{(a+3)} \\[\n]
	C_{uB}^{(3a)*}	\\[\n]
	C_{uW}^{(3a)*} \\[\n]
	C_{uG}^{(3a)*}
\end{array}\right)
\\
+&\us{0.082}{+1} \left( |C_{lequ}^{1(13)}|^2 + |C_{lequ}^{1(31)}|^2 \right)
+ \us{3.5  }{-8} \left( |C_{lequ}^{3(13)}|^2 + |C_{lequ}^{3(31)}|^2 \right)
\\[5mm]
\Gamma&_{t\to j\,\ell^+\ell^-}^\text{on-peak}/10^{-5}\,\text{GeV}
\;\times (\Lambda/1\,\text{TeV})^4 =
\\
&\Re
\left(\begin{array}{@{}l@{}}
	C_{lq}^{-(a+3)}  \\[\n]
	C_{eq}^{(a+3)} \\[\n]
	C_{\varphi q}^{-(a+3)} \\[\n]
	C_{uB}^{(a3)}	\\[\n]
	C_{uW}^{(a3)} \\[\n]
	C_{uG}^{(a3)}
\end{array}\right)^{\hspace{-1.5mm}\dagger}
\left(
\begin{array}{*{5}{@{}c@{\hspace{3mm}}}c@{}} 
+\us{0.069}{-9}
	& 	0
	& -\us{0.02}{+6}	-\us{0.2}{-9}i
	& -\us{0.053}{-5}	-\us{0.1}{-8}i
	& -\us{0.052}{-16}	+\us{0.34}{-8}i
	& +\usv{0.014}{---}	-\usv{0.013}{---}i 
\\	
	& +\us{0.069}{-9}
	& +\us{0.017}{+6}	+\us{0.18}{-9}i
	& -\us{0.053}{-10}	+\us{0.09}{-8}i
	& -\us{0.054}{+0}	-\us{0.3}{-8}i
	& -\usv{0.007}{---}	+\usv{0.017}{---}i 
\\	
	& 	
	& +\us{1.7}{-9}
	& +\us{1.7}{-8}	-\us{0.0095}{-8}i
	& -\us{5.7}{-8}	-\us{0.0095}{-8}i
	& +\usv{0.27}{---}	+\usv{0.2}{---}i 
\\	
	& 	
	& 	
	& +\us{0.64}{-9}
	& -\us{3.9}{-9}	-\us{0.029}{-9}i
	& +\usv{0.16}{---}	+\usv{0.14}{---}i 
\\	
	& 	
	& 	
	& 	
	& +\us{6.6}{-9}
	& -\usv{0.53}{---}	-\usv{0.47}{---}i 
\\	
	& 	
	& 	
	& 	
	& 	
	& +\usv{0.002}{---}
\end{array}\right)
\left(\begin{array}{@{}l@{}}
	C_{lq}^{-(a+3)}  \\[\n]
	C_{eq}^{(a+3)} \\[\n]
	C_{\varphi q}^{-(a+3)} \\[\n]
	C_{uB}^{(a3)}	\\[\n]
	C_{uW}^{(a3)} \\[\n]
	C_{uG}^{(a3)}
\end{array}\right)
\\
+& \Re	
\left(\begin{array}{@{}l@{}}
	C_{lu}^{(a+3)}  \\[\n]
	C_{eu}^{(a+3)} \\[\n]
	C_{\varphi u}^{(a+3)} \\[\n]
	C_{uB}^{(3a)*}	\\[\n]
	C_{uW}^{(3a)*} \\[\n]
	C_{uG}^{(3a)*}
\end{array}\right)^{\hspace{-1.5mm}\dagger}
\left(
\begin{array}{*{5}{@{}c@{\hspace{3mm}}}c@{}}
+\us{0.069}{-9}
	& 	0
	& -\us{0.02}{+6}	-\us{0.2}{-9}i
	& -\us{0.053}{-5}	-\us{0.1}{-8}i
	& -\us{0.052}{-16}	+\us{0.34}{-8}i
	& -\usv{0.002}{---}	+\usv{0.013}{---}i 
\\	
	& +\us{0.069}{-9}
	& +\us{0.017}{+6}	+\us{0.18}{-9}i
	& -\us{0.053}{-10}	+\us{0.09}{-8}i
	& -\us{0.054}{+0}	-\us{0.3}{-8}i
	& +\usv{0.0067}{---}	-\usv{0.006}{---}i 
\\	
	& 	
	& +\us{1.7}{-9}
	& +\us{1.7}{-8}	-\us{0.0095}{-8}i
	& -\us{5.7}{-8}	-\us{0.0095}{-8}i
	& -\usv{0.17}{---}	-\usv{0.09}{---}i 
\\	
	& 	
	& 	
	& +\us{0.64}{-9}
	& -\us{3.9}{-9}	-\us{0.029}{-9}i
	& -\usv{0.098}{---}	-\usv{0.068}{---}i 
\\	
	& 	
	& 	
	& 	
	& +\us{6.6}{-9}
	& +\usv{0.31}{---}	+\usv{0.21}{---}i 
\\	
	& 	
	& 	
	& 	
	& 	
	& +\usv{0.00066}{---} 
\end{array}\right)
\left(\begin{array}{@{}l@{}}
	C_{lu}^{(a+3)}  \\[\n]
	C_{eu}^{(a+3)} \\[\n]
	C_{\varphi u}^{(a+3)} \\[\n]
	C_{uB}^{(3a)*}	\\[\n]
	C_{uW}^{(3a)*} \\[\n]
	C_{uG}^{(3a)*}
\end{array}\right)
\\
+&\us{0.02}{0} \left( |C_{lequ}^{1(13)}|^2 + |C_{lequ}^{1(31)}|^2 \right)
+ \us{0.81}{-9}  \left( |C_{lequ}^{3(13)}|^2 + |C_{lequ}^{3(31)}|^2 \right)
\end{align*}
\caption{Contributions of each FCNC operator coefficient to the $t\to
j\:\ell^+\ell^-$ partial width (for one single species of massless quark and
charged leptons)~\cite{Zhang:2014rja}. The subscripts indicate the relative
correction brought by the NLO contribution in QCD (a dash stresses the absence
of leading-order contribution). Off-shell $Z$ effects are included to all
orders. In the \emph{on-peak\,+\,off-peak} case, a $15$~GeV cut has been applied
on the invariant mass of the two leptons, to avoid the divergence of the $t\to
j\gamma^*\to j\,\ell^+\ell^-$ contribution. In the \emph{on-peak} case,
$m_{\ell\ell}\in[78,102]$~GeV is required.
}
\label{tab:t_to_uee}
\end{table*}

\begin{figure}[tb]\centering
\includegraphics[width=\columnwidth]{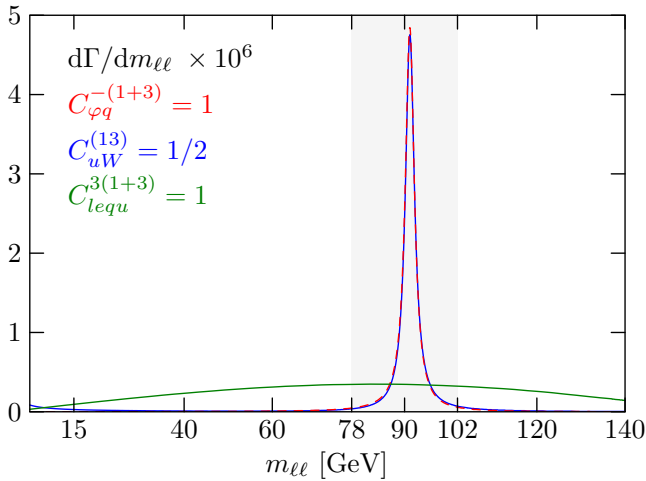}
\caption{Invariant mass distribution of lepton pair in $t\to j\,\ell^+\ell^-$.
Contributions from two-quark $O_{\varphi q}^{-}$ and $O_{uW}$ as
well as two-quark--two-lepton $O_{lequ}^{3}$ operators are compared.}
\label{fig:4fdecay}
\end{figure}

Remarkably, for operator coefficients of equal magnitude, the
$O_{lequ}^{3}$ two-quark--two-lepton operators contribute in proportions
comparable to two-quark ones. The lepton invariant mass distributions of
two-quark operator contributions are however strongly peaked around $m_Z$
(see \autoref{fig:4fdecay}). 
Current searches for top-quark FCNC decays to a light jet and a lepton pair
actually focus on lepton invariant mass windows close to the $Z$-boson mass
(\emph{e.g.}, $m_{\ell\ell}\in[78,102]$~GeV) and interpret the obtained result
as limits on $t\to jZ$, without taking into account two-quark--two-lepton
operator contributions. However, even in this on-$Z$-peak window, the second
part of \autoref{tab:t_to_uee} shows some residual sensitivity to
two-quark--two-lepton operators, $O_{lequ}^{3}$ especially. Neglecting
interferences and considering only operators $O_{\varphi q}^{-}$, $O_{uW}$ and
$O_{lequ}^{3}$ for the sake of illustration, \autoref{tab:t_to_uee} gives:
\begin{multline}
  \Gamma^\text{on-peak}_{t\to u\:e^+e^-}
  \;/10^{-5}\ \mathrm{GeV}
  \;\times (\Lambda/1\,\text{TeV})^4\\*
  =
     1.7\:|{C_{\varphi q}^{-(1+3)}}|^2
    +6.6\:|{C_{uW}^{(13)}}|^2
    +0.81\:|{C_{lequ}^{3(13)}}|^2
    \label{eq:onshellrate}
\end{multline}
On the other hand, in the off-peak region of the spectrum, $m_{\ell\ell} \in
[15,78] \cup [102,\infty]$~GeV, one has:
\begin{multline}
  \Gamma^\text{off-peak}_{t\to u\:e^+e^-}
  \;/10^{-5}\ \mathrm{GeV}
  \;\times (\Lambda/1\,\text{TeV})^4\\*
  =
   0.2\:|C_{\varphi q}^{-(1+3)}|^2
  +1.0\:|C_{uW}^{(13)}|^2
  +2.7\:|C_{lequ}^{3(13)}|^2
    \label{eq:offshellrate}
\end{multline}
By distinguishing both regions, one therefore gets a means of constraining
separately two-quark and two-quark--two-lepton operators. Moreover, were a signal to be
observed, its proportion in each of those ranges of lepton invariant masses
would bring information about its nature. As the off-peak region contains less
Drell-Yan background, a better sensitivity may actually be obtained on
two-quark--two-lepton operators coefficients.

Similarly, one may use angular distributions to disentangle the contributions
of vector, scalar and tensor operators, as done in Ref.~\cite{Aaltonen:2009ef}.
In Ref.~\cite{Zhang:2014rja}, the $Z$ helicity fractions were also computed at
NLO in QCD as functions of operators coefficients. Taking into account
differential decay rates should therefore allow to disentangle all types of
operators. Since identifying final-state up- and charm-quark jets can only be done
with a limited efficiency, one should rely on production processes and take
benefit of the widely different parton distribution functions of $u$'s and $c$'s
to discriminate between both contributions.

\subsection{Single-top production}\label{sec:production}
\vspace*{-2mm}

\begin{figure}[tb]\centering
	\includegraphics[width=.82\linewidth]{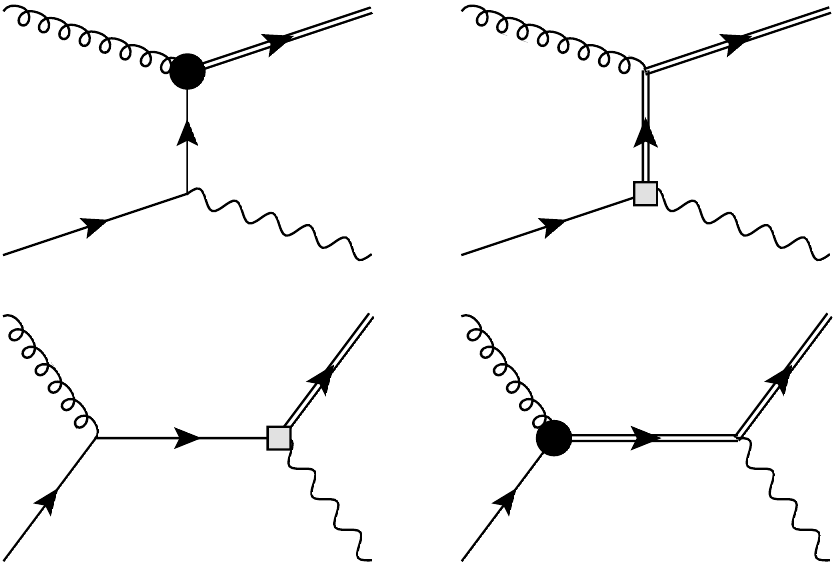}
	\caption{Tree-level diagrams for $pp\to t\gamma$ and $pp\to tZ$.}
	\label{fig:pptv}
\end{figure}

\begin{figure}[tb]\centering
	\includegraphics[width=.82\linewidth]{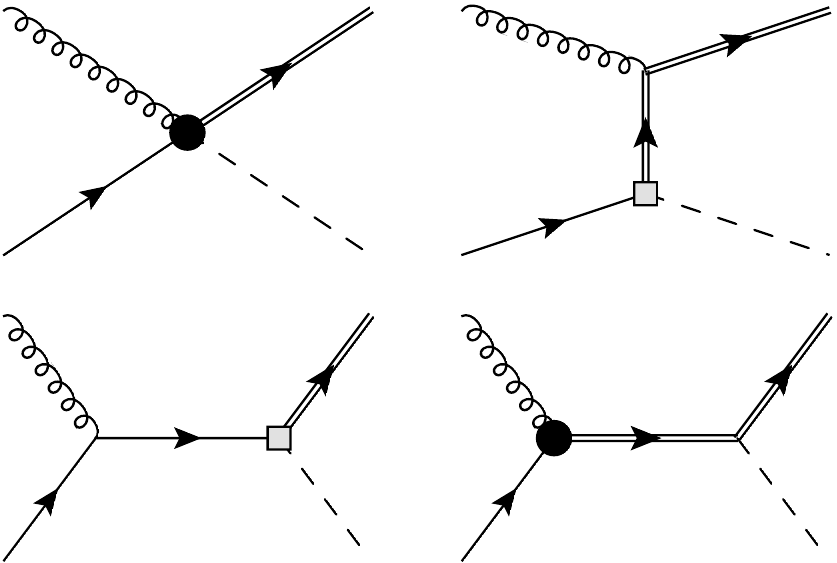}
	\caption{Tree-level diagrams for $pp\to th$.}
	\label{fig:ppth}
\end{figure}

Single-top production associated with a neutral gauge boson, $\gamma$, $Z$, or
the scalar boson $h$ can bring useful information on
top-quark FCNC \cite{AguilarSaavedra:2004wm, AguilarSaavedra:2000aj,
Atwood:2013ica, Kao:2011aa, Agram:2013koa, Khatibi:2014via, Greljo:2014dka,
Kidonakis:2003sc, Khanpour:2014xla}. As illustrated in \autoref{fig:pptv} and
\autoref{fig:ppth},
$O_{uG}$ already contributes at leading order to $pp\to t\gamma$ or $pp\to tZ$
and $pp\to th$. Following the general strategy outlined in
Ref.~\cite{Christensen:2009jx}, all two-quark operators have been implemented in
the {\sc FeynRules}/{\sc MadGraph5\_aMC@NLO} simulation
chain~\cite{Alloul:2013bka,Degrande:2011ua,Alwall:2014hca} which permits
automated NLO computations in QCD, including matching to parton shower. The
details of this implementation are discussed elsewhere~\cite{Degrande:2014tta}.


\begin{figure}[tb]
\vspace{-3mm}
\includegraphics[width=\columnwidth]{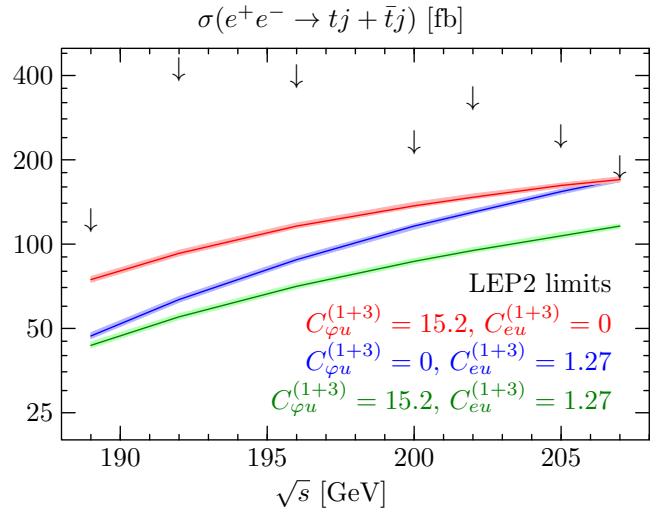}
\caption{Cross section [fb] for $e^+e^-\to tj+\bar tj$ for three illustrative
choices of parameters at NLO accuracy in QCD (lines); $95\%$ CL limits (arrows)
set by a combination ALEPH, DELPHI, L3 and OPAL results~\cite{LEP2001}. The
uncertainty bands (${}^{+2.2\%}_{-1.8\%}$ at $\sqrt{s}=207$~GeV) are obtained by
running $\alpha_s$ from $m_t/2$ to $2m_t$ as the anomalous dimensions of vector
operators vanish.}
\label{fig:ee_tu}
\end{figure}

\begin{figure}[tb]
\includegraphics[width=\columnwidth]{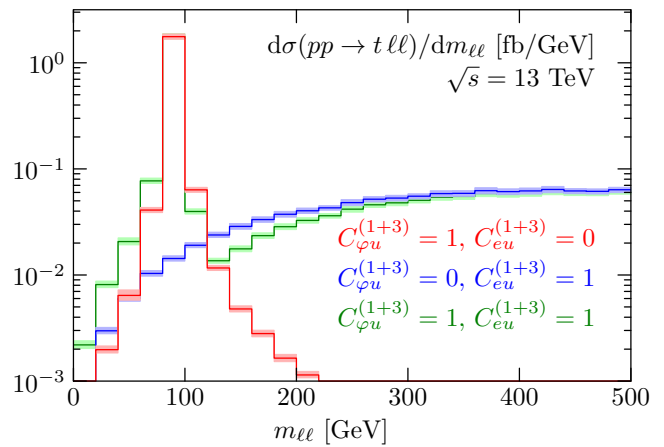}
\caption{NLO lepton invariant mass distribution in $pp\to t\ell\ell$, from the two-quark operator $O_{\varphi u}$ only (red), from the two-quark--two-lepton operators $O_{eu}$ (blue) and from their interference (green).}
\label{fig:pp2tll}
\end{figure}

A $m_{\ell\ell}$-dependent reweighing can also be used to obtain, from
two-quark operator results, the NLO-accurate contributions of vector (and
scalar) two-quark--two-lepton operators. Such operators have not yet been
implemented in {\sc MadGraph5\_aMC@NLO}. \autoref{fig:ee_tu} compares $e^+e^-\to
tj+\bar tj$ production cross sections through a two-quark and a
two-quark--two-lepton operator, as well as through their interference. The bounds
deriving from a combination of LEP2 results~\cite{LEP2001} are also shown. In
\autoref{fig:pp2tll} the contributions of the same two operators, $O_{\varphi u}$
and $O_{eu}$, to $pp\to t\,\ell^+\ell^-$ at $\sqrt{s}=13$\,~TeV are shown. In
those figures, the uncertainty bands are obtained from factorization and
renormalization scale variations between $m_t/2$ and $2m_t$.

\section{A first global analysis}\label{sec:global}
\vspace*{-2mm}

In this section we illustrate the feasibility of a global approach to top-quark
FCNC interactions. For the sake of illustration and simplicity, we only consider
the most constraining observables. This suffices to set significant bounds on
all two-quark operators listed previously as well as on a subset of the
two-quark--two-lepton ones. Four-fermion operators featuring two leptons of
different generations or a pair of taus would remain unconstrained
due to the absence of experimental searches while those with
two muons are only loosely bound due to the lack of off-$Z$-peak constraint in
$t\to j\,\ell^+\ell^-$ searches. LEP2 data, however, effectively constrains
operators containing an electron pair. We will neglect the contributions of
four-quark operators to considered observables. They are suppressed by an
imposed jet veto in $pp\to t$ and only appear at NLO in QCD in the other
processes we take into account.

Currently, for either $j=u$ or $c$, the most constraining $95\%$ CL bounds on
the top-quark branching ratios are:
\begin{align*}
	\Br(t\to j\,e^+e^-) + \Br(t\to j\,\mu^+\mu^-)\lesssim 0.0017\%
	\quad	&\mbox{\cite{Chatrchyan:2013nwa},\footnotemark}
\\
	\Br(t\to j\gamma)<3.2\%
	\quad	&\mbox{\cite{Abe:1997fz},\;\;}
\\
	\Br(t\to j\,\gamma\gamma)<0.0016\%
	\quad	&\mbox{\cite{CMS:2014qxa}.\footnotemark}
\end{align*}
\footnotetext[2]{Those two figures are obtained using $\Br(Z\to \ell^+\ell^-) =
3.37\%$~\cite{Beringer:1900zz} as well as the CMS limit on $\Br(t\to
jZ)<0.05\%$ which combines the $e^+e^-$ and $\mu^+\mu^-$ channels. They may therefore be
slightly underestimated and do not account for the difference in efficiency of
these two channels.}%
\footnotetext[3]{The limit on $\Br(t\to ch)<0.69\%$ and the assumed
$\Br(h\to\gamma\gamma)=0.23\%$ both quoted in Table~4 have been used.}%
(Top and antitop branching fractions are assumed identical by
the experimental collaborations.) 
The first limit is actually applicable for lepton invariant masses close to
$m_Z$: $m_{\ell\ell}\in[78,102]$~GeV. The CMS Collaboration actually
interprets it as a bound on $\Br(t\to j\,Z)$ even though
four-fermion operator contributions cannot in general be neglected. Similarly, the third
limit is obtained for $m_{\gamma\gamma}\in [120,130]$~GeV and interpreted as
a bound on $\Br(t\to ch)$ while the contributions of $uth$ and $u(c)\,t\gamma$
interactions should in principle also be taken into account.
Furthermore, a limit on the single-top production cross section~\cite{TheATLAScollaboration:2013vha}:
\begin{align*}
	\sigma( pp\to t )+\sigma( pp\to \bar t ) <2.5~\text{pb \quad
	at }\sqrt{s}=8~\text{TeV}
\end{align*}
is converted by the ATLAS Collaboration into the $\Br(t\to ug)<0.0031\%$ and
$\Br(t\to cg)<0.016\%$ bounds on top-quark FCNC branching fractions when a
$tug$ or a $tcg$ vertex are respectively assumed to provide the only
contributions to the above cross section.
Similarly, the
\begin{multline*}
	\sigma(ug\to t\gamma ) + \sigma( ug\to \bar t\gamma)
	 \\+ 0.778\: \left[
	\sigma(cg\to t\gamma) + \sigma(cg\to \bar t\gamma)
	\right]\\ < 0.0670~\text{pb\quad
	at }\sqrt{s_{pp}}=8~\text{TeV},
\end{multline*}
bound obtained by the CMS Collaboration~\cite{CMS:2014hwa}\footnote{This
expression is obtained from the bounds on the NLO cross sections times $W$
leptonic branching fraction (we took $\Br(W\to l\nu_l) = 3\times
10.80\%$~\cite{Beringer:1900zz}) provided by the CMS Collaboration.}
for $p_{T\gamma}>30$~GeV
is translated into the $\Br(t\to u\gamma)<0.0108\%$ and $\Br(t\to c\gamma)<0.132\%$
limits by taking into account either $ut\gamma$ or $ct\gamma$ contributions only
(the $utg$ and $ctg$ contributions are notably assumed vanishing). Finally, a
LEP2 combination~\cite{LEP2001} implies
\begin{gather*}
\sigma(e^+e^-\to tj+\bar tj) < 176~\text{fb}
\qquad\text{at }\sqrt{s}=207\,\text{GeV}
\end{gather*}
for $m_t=172.5$~GeV.

\subsection{\texorpdfstring{$\boldsymbol{t\to j\,\ell^+\ell^-}$}{t > j l l}}
\vspace*{-2mm}
Let us first consider the top decay to a pair of charged leptons and a jet.
It is mainly sensitive to operators inducing a $t\to jZ$ decay. At
leading order, the rate for this process can be expressed as a sum of four
squared terms corresponding to final states of different polarizations
($q_LZ_0$, $q_RZ_0$, $q_LZ_-$ and $q_RZ_+$):
\begin{align*}
  \Gamma_{t\to jZ}=&\frac{\alpha m_t^5 (1-x^2)^2}{8\Lambda^4s_W^2c_W^2}
  \\
  \sum_{a=1,2}
  \Big\{
  &\:\Big|
    \frac{1}{2x}C_{\varphi q}^{-(a+3)}
    +2x\Big(s_W^2C_{uB}^{(a3)}-c_W^2C_{uW}^{(a3)}\Big)
    \Big|^2
  \nonumber\\[-2mm]
  +&\:\Big|
    \frac{1}{2x}C_{\varphi u}^{(a+3)}
    +2x\Big(s_W^2C_{uB}^{(3a)*}-c_W^2C_{uW}^{(3a)*}\Big)
    \Big|^2
  \nonumber\\
  +2&\:\Big| \frac{1}{2}C_{\varphi q}^{-(a+3)}
  +2\Big(s_W^2C_{uB}^{(a3)} - c_W^2C_{uW}^{(a3)} \Big) \Big|^2
  \nonumber\\
  +2&\:\Big| \frac{1}{2}C_{\varphi u}^{(a+3)}
  +2\Big(s_W^2C_{uB}^{(3a)*} - c_W^2C_{uW}^{(3a)*} \Big) \Big|^2
  \Big\},
\end{align*}
where $x\equiv m_Z/m_t$, $s_W\equiv\sin\theta_W$ and $c_W\equiv\cos\theta_W$.
Therefore, $t\to j\,\ell^+\ell^-$ in the on-$Z$-peak region dominantly
constrains four linear combinations of $C^{-(a+3)}_{\varphi
q},C^{(a+3)}_{\varphi u}$, \; $C_{uB}^{(a3)},C_{uW}^{(a3)}$ and
$C_{uB}^{(3a)},C_{uW}^{(3a)}$ for both $a=1$ and $2$.

Numerical results that are NLO accurate in QCD, include the full $\Gamma_Z$
dependence and all the two-quark--two-lepton operators have been collected in
\autoref{tab:t_to_uee}. At that order, a dependence on the $O_{uG}$ operator
coefficients is generated. It has, however, little overall effect given the tight
constraints of $C_{uG}$ that arise from $pp\to t,\:\bar t$ searches.

\subsection{\texorpdfstring{$\boldsymbol{pp\to t,\;\bar t}$}{pp > t, tbar}}\label{sec:gluon}
\vspace*{-2mm}

The most sensitive of the single-top production limits constrains the $C_{uG}$
coefficients alone, provided the four-quark operator contributions in the
experimental acceptance are neglected. Using the NLO result:
\begin{equation*}
\Gamma_{t\to jj}= B\:\Gamma_t  \left(\frac{1\,\text{TeV}}{\Lambda}\right)^4 \sum_{a=1,2}\Big(\:
	|C_{uG}^{(a3)}|^2 + 
	|C_{uG}^{(3a)}|^2 
	\Big),
\end{equation*}
with $B\equiv0.0186$ and a fixed value for the top width
$\Gamma_t=1.32$~GeV, we recast the interpretation made in
Ref.~\cite{TheATLAScollaboration:2013vha} to obtain the
bound on the operator coefficient combination actually probed in $pp\to t+\bar t$:
\begin{multline}
\frac{1}{B_u} ( 
	\big|C_{uG}^{(13)}\big|^2 +
	\big|C_{uG}^{(31)}\big|^2
) \\[-3mm]+ \frac{1}{B_c} (
	\big|C_{uG}^{(23)}\big|^2 +
	\big|C_{uG}^{(32)}\big|^2
)
< \frac{1}{B} \left(\frac{\Lambda}{1\,\text{TeV}}\right)^4,
\nonumber
\end{multline}
where $Br(t\to ug)<0.0031\%\equiv B_u$ and $\Br(t\to cg)<0.016\%\equiv B_c$ are
the limits set assuming one single contribution from either $a=1$ or $a=2$.
Fixing $\Lambda=1\,$TeV, the following ---strong--- constraints are obtained on the
coefficient moduli:
\begin{center}
	\includegraphics[scale=1]{limits_0}
\end{center}
where the red allowed range applies to up-quark operator coefficients ($a=1$)
and blue ranges to charm-quark ones ($a=2$).

Actually, at this stage, all operator coefficients but the $C_{u\varphi}$ ones
are already constrained, sometimes poorly though. Most notably, the $t\to
j\,\ell^+\ell^-$ observable is primarily sensitive to the $s_W^2
O_{uB}-c_W^2O_{uW}$ linear combinations that contain the tensorial $Z$
interactions, while $O_{uB}+O_{uW}$ contains the photon ones. The absence of
experimental bound outside the on-$Z$-peak region for $m_{\ell\ell}$ also
renders the two-quark--two-lepton operators very loosely constrained.
Quantitatively with the two observables just described, the limits that arise on
the moduli of the operator coefficients are:
\begin{center}
	\includegraphics[scale=1]{limits_1}
\end{center}
where the white marks indicate the bounds that would have been obtained out of
our global picture, by assuming that all coefficients but the constrained one vanish.
Among those limits, only the ones applying to $|C^-_{\varphi q}|$ or $|C_{\varphi
u}|$ will not improve much in what follows.

\subsection{\texorpdfstring{$\boldsymbol{t\to j\gamma}\quad$ and $\quad\boldsymbol{pp\to t\gamma,\;\bar t\gamma}$}{t > j a  and  pp > t a, tbar a}}\label{sec:photon}
\vspace*{-2mm}

Breaking the approximate degeneracy between the $C_{uB}$ and $C_{uW}$
coefficients present in $t\to j\,\ell^+\ell^-$ can be achieved through top-quark FCNC
processes involving a photon. The CDF Collaboration still currently sets the best
limit on the top decay to a photon and a jet: $\Br(t\to j
\gamma)<3.2\%$~\cite{Abe:1997fz}. The corresponding leading-order decay rate
writes:
\begin{equation*}
  \Gamma_{t\to j\gamma}=\frac{\alpha m_t^5}{\Lambda^4}\sum_{a=1,2}\Big( 
    \big| C_{uB}^{(a3)}+C_{uW}^{(a3)} \big|^2
   +\big| C_{uB}^{(3a)}+C_{uW}^{(3a)} \big|^2\Big).
\end{equation*}
At order $\mathcal{O}(\alpha_s)$, a cut on the photon energy and jet-photon
separation is required to avoid the soft-collinear divergence. For
$(E_{\gamma,j},\mathbf{p}_{\gamma,j})$ the quadrimomenta of the photon and
quark jet in the top rest frame, we take:
\begin{align*}
  1-\mathbf{p}_\gamma\cdot\mathbf{p}_{j}/E_\gamma E_j>0.2,\\
  E_\gamma>20\ \mathrm{GeV}.
\end{align*}
With these cuts the CDF constraint reads
\begin{align}
 \sum_{a=1,2}\Big\{
  &\big|0.71\:C_{uB}^{(a3)}+0.71\:C_{uW}^{(a3)}-0.036\:C_{uG}^{(a3)}\big|^2
  \nonumber\\[-4mm]
  &\big|0.71\:C_{uB}^{(3a)}+0.71\:C_{uW}^{(3a)}-0.036\:C_{uG}^{(3a)}\big|^2
  \Big\}
  \nonumber\\
  <\:& 19.6\:  \frac{3.2\%}{\Br_{t\to j\gamma}^\text{exp}} \frac{\Gamma_t}{1.32\,\text{GeV}} \bigg(\frac{\Lambda}{1\,\text{TeV}}\bigg)^4.
\label{eq:aj_bound}
\end{align}

\begin{figure}[tb]
\includegraphics[width=\columnwidth]{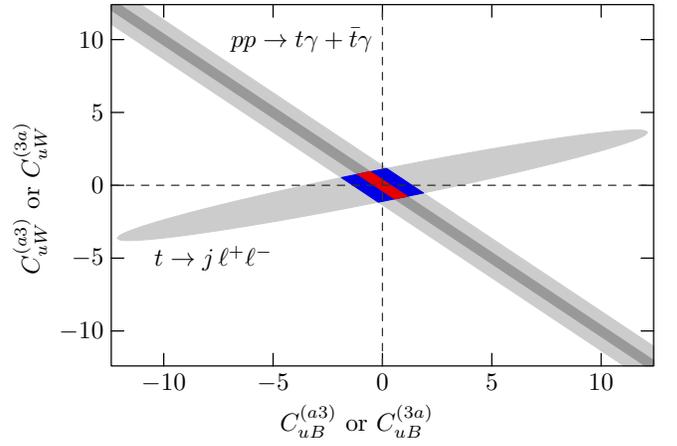}
\caption{Complementarity of the $t\to j\,\ell^+\ell^-$ and $pp\to t\gamma,\,\bar
t\gamma$ limits in the $C_{uB}-C_{uW}$ plane. The $C_{uG}$ coefficients are
constrained to satisfy the bounds set by the $pp\to t,\bar t$ searches. The dark
gray and red allowed regions apply for $a=1$ while the blue intersection shows the
constraint for $a=2$. The same limits apply to either the real or the imaginary
parts of the operator coefficients.}
\label{fig:limits_uBuW}
\end{figure}
A much stronger constraint on $C_{uB}+C_{uW}$ is actually obtained by
considering the bound set in Ref.~\cite{CMS:2014hwa} on single-top production in
association with a photon of transverse momentum $p_{T\gamma}>30$~GeV. Taking
into account the relative efficiency obtained by CMS for up-gluon and
charm-gluon initial states and NLO results in QCD for $\sigma(pp\to \:t\gamma+\bar
t\gamma)$ at $\sqrt{s}=8$~TeV obtained with the implementation of
Ref.~\cite{Degrande:2014tta} in aMC@NLO~\cite{Alwall:2014hca} we get:
\begin{align*}
&
\left(\begin{array}{@{}l@{}}
	C_{uB}^{(13)}	\\[\n]
	C_{uW}^{(13)} \\[\n]
	C_{uG}^{(13)}
\end{array}\right)^{\hspace{-1.5mm}\dagger}
\left(
\begin{array}{*{2}{@{}c@{\hspace{3mm}}}c}
	  \us{0.46}{}
	& \us{0.93}{}
	& \us{0.2}{} 
\\	
	& \us{0.46}{}
	& \us{0.2}{} 
\\	
	& 	
	& \us{1.9}{}
\end{array}\right)
\left(\begin{array}{@{}l@{}}
	C_{uB}^{(13)}	\\[\n]
	C_{uW}^{(13)} \\[\n]
	C_{uG}^{(13)}
\end{array}\right)
\\+ (13)&\leftrightarrow(31)
\\
+0.78&
\left(\begin{array}{@{}l@{}}
	C_{uB}^{(23)}	\\[\n]
	C_{uW}^{(23)} \\[\n]
	C_{uG}^{(23)}
\end{array}\right)^{\hspace{-1.5mm}\dagger}
\left(
\begin{array}{*{2}{@{}c@{\hspace{3mm}}}c}
	  \us{0.047}{}
	& \us{0.095}{}
	& \us{0.017}{} 
\\	
	& \us{0.047}{}
	& \us{0.017}{} 
\\	
	& 	
	& \us{0.33}{} 
\end{array}\right)
\left(\begin{array}{@{}l@{}}
	C_{uB}^{(23)}	\\[\n]
	C_{uW}^{(23)} \\[\n]
	C_{uG}^{(23)}
\end{array}\right)
\\+ (23)&\leftrightarrow(32)
\\&<0.067\quad (\Lambda / 1\,\text{TeV})^4
\end{align*}
With this observable taken into account, the limits on $C_{uB}$ and $C_{uW}$
improve dramatically. Indirectly, the bounds on two-quark--two-lepton operators
that interfere with those in $t\to j\,\ell^+\ell^-$ also improve slightly:
\begin{center}
	\includegraphics[scale=1]{limits_2}
\end{center}
The complementarity of $t\to j\,\ell^+\ell^-$ and $pp\to t\gamma,\,\bar t\gamma$
observables is illustrated in \autoref{fig:limits_uBuW}.

\begin{table*}[htb]
\begin{align*}
\sigma&_{e^+e^-\to tj}^{\sqrt{s}=207\,\text{GeV}}\,\text{[fb]}
\;\times (\Lambda/1\,\text{TeV})^4 =
\\
&\Re
\left(\begin{array}{@{}l@{}}
	C_{lq}^{-(a+3)*}  \\[\n]
	C_{eq}^{(a+3)*} \\[\n]
	C_{\varphi q}^{-(1+3)*} \\[\n]
	C_{uB}^{(a3)*}	\\[\n]
	C_{uW}^{(a3)*} \\[\n]
	C_{uG}^{(a3)*}
\end{array}\right)^{\hspace{-1.5mm}\dagger}
\left(
\begin{array}{*{5}{@{}c@{\hspace{3mm}}}c}
\us{+52}{+24}
	& 	0
	& \us{+6.5}{+25}	\us{-0.035}{}i
	& \us{-9}{+24}	\us{-0.036}{}i
	& \us{-38}{+24}	\us{+0.12}{}i
	& \usv{+1}{---} 
\\	
	& \us{+52}{+24}
	& \us{-5.8}{+25}	\us{+0.03}{}i
	& \us{-22}{+24}	\us{+0.032}{}i
	& \us{+3.8}{+25}	\us{-0.1}{}i
	& \usv{+0.04}{---} 
\\	
	& 	
	& \us{+0.37}{+25}
	& \us{+0.63}{+24}	\us{-0.00064}{}i
	& \us{-2.6}{+25}	\us{-0.00064}{}i
	& \usv{+0.061}{---} 
\\	
	& 	
	& 	
	& \us{+2.7}{+25}
	& \us{+2.5}{+23}	\us{-0.003}{}i
	& \usv{-0.1}{---} 
\\	
	& 	
	& 	
	& 	
	& \us{+7.3}{+25}
	& \usv{-0.37}{---} 
\\	
	& 	
	& 	
	& 	
	& 	
	& \usv{+1.6\times 10^{-5}}{---} 
\end{array}\right)
\left(\begin{array}{@{}l@{}}
	C_{lq}^{-(a+3)*}  \\[\n]
	C_{eq}^{(a+3)*} \\[\n]
	C_{\varphi q}^{-(a+3)*} \\[\n]
	C_{uB}^{(a3)*}	\\[\n]
	C_{uW}^{(a3)*} \\[\n]
	C_{uG}^{(a3)*}
\end{array}\right)
\\
+& \Re	
\left(\begin{array}{@{}l@{}}
	C_{lu}^{(a+3)*}  \\[\n]
	C_{eu}^{(a+3)*} \\[\n]
	C_{\varphi u}^{(a+3)*} \\[\n]
	C_{uB}^{(3a)}	\\[\n]
	C_{uW}^{(3a)} \\[\n]
	C_{uG}^{(3a)}
\end{array}\right)^{\hspace{-1.5mm}\dagger}
\left(
\begin{array}{*{5}{@{}c@{\hspace{3mm}}}c}
\us{+52}{+24}
	& 	0
	& \us{+6.5}{+25}	\us{-0.035}{}i
	& \us{-9}{+24}	\us{-0.036}{}i
	& \us{-38}{+24}	\us{+0.12}{}i
	& \usv{+1}{---} 
\\	
	& \us{+52}{+24}
	& \us{-5.7}{+24}	\us{+0.03}{}i
	& \us{-22}{+24}	\us{+0.032}{}i
	& \us{+3.8}{+25}	\us{-0.1}{}i
	& \usv{+0.71}{---} 
\\	
	& 	
	& \us{+0.37}{+25}
	& \us{+0.63}{+24}	\us{-0.00064}{}i
	& \us{-2.6}{+25}	\us{-0.00064}{}i
	& \usv{+0.024}{---} 
\\	
	& 	
	& 	
	& \us{+2.7}{+24}
	& \us{+2.5}{+23}	\us{-0.003}{}i
	& \usv{-0.24}{---} 
\\	
	& 	
	& 	
	& 	
	& \us{+7.3}{+25}
	& \usv{-0.35}{---} 
\\	
	& 	
	& 	
	& 	
	& 	
	& \usv{+1.6\times 10^{-5}}{---} 
\end{array}\right)
\left(\begin{array}{@{}l@{}}
	C_{lu}^{(a+3)*}  \\[\n]
	C_{eu}^{(a+3)*} \\[\n]
	C_{\varphi u}^{(1+3)*} \\[\n]
	C_{uB}^{(3a)}	\\[\n]
	C_{uW}^{(3a)} \\[\n]
	C_{uG}^{(3a)}
\end{array}\right)
\\
+&\us{33}{+42} \left( |C_{lequ}^{1(a3)}|^2 + |C_{lequ}^{1(3a)}|^2 \right)
+ \us{370}{+26} \left( |C_{lequ}^{3(a3)}|^2 + |C_{lequ}^{3(3a)}|^2 \right)
\end{align*}

\caption{NLO expression for the $e^+e^-\to tj$ cross section in fb, including
full $\Gamma_Z$ dependence, for either $a=1$ or $2$ (light quark and electron
masses are neglected). For $e^+e^-\to \bar tj$ the two complex matrices should
be conjugated. As the experimental constraint is set on $tj$ plus $\bar tj$
production, the imaginary prefactors provided here at leading order only have no
effect on our limits. Identical bounds therefore apply to the real and imaginary
parts of each operator coefficient. The $C_{lequ}^{1,3}$ prefactors are
obtained fully analytically (see Appendix\,\ref{sec:k-fac}) while the other
ones, from the aMC@NLO implementation of Ref.~\cite{Degrande:2014tta} either
directly, for two-fermion operators, or after reweighing, for four-fermion
ones.}
\label{tab:ee2tq}
\vspace*{-3mm}
\end{table*}

\vspace*{-2mm}
\subsection{\texorpdfstring{$\boldsymbol{e^+e^-\to tj,\;\bar tj}$}{e e > t j, tbar j}}
\vspace*{-2mm}
Without an off-$Z$-peak constraint on $t\to j\,\ell^+\ell^-$, one cannot do
better for the two-quark--two-lepton operators involving muons. However, the
limit set at LEP on $e^+e^-\to tj+\bar tj$~\cite{LEP2001} gives a powerful handle on
the two-quark--two-lepton operators involving electrons. The next-to-leading-order
expression for $\sigma(e^+e^-\to tj)$ at $\sqrt{s}=207$~GeV is provided in
\autoref{tab:ee2tq}. As can be seen from \autoref{fig:ee_tu}, the bound set at
this center-of-mass energy is the most constraining one. We do not attempt a
combination of the bounds set at different center-of-mass energies that could
only be naive given the lack of published statistical information.

\begin{figure}[tb]
\includegraphics[width=\columnwidth]{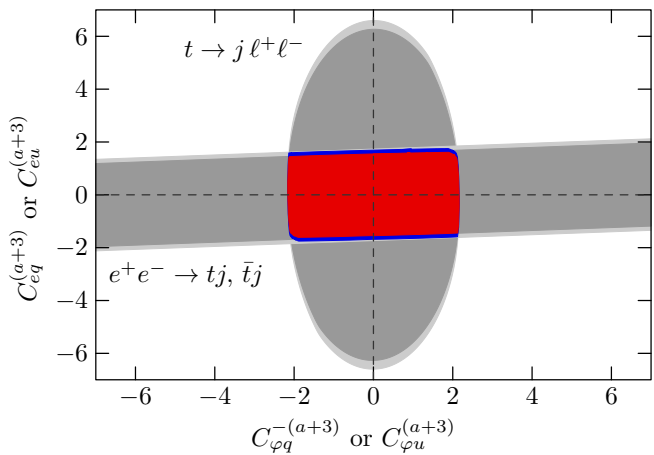}
\caption{Complementarity of the $e^+e^-\to tj+\bar tj$ and $t\to j\,e^+e^-$
limits for constraining two- and four-fermion operators. The operator
coefficients not shown in this plane are constrained to satisfy the bound of
\autoref{fig:limits}. The dark gray and red allowed regions apply for $a=1$
while the light gray and blue ones for $a=2$. The same limits apply to either
the real or the imaginary parts of the operator coefficients.}
\label{fig:limits_pu_eu}
\end{figure}

The complementarity of the LEP limit with the $t\to j\,e^+e^-$ one for
constraining simultaneously two- and four-fermion operators is illustrated in
\autoref{fig:limits_pu_eu}.

\vspace*{-3mm}
\subsection{\texorpdfstring{$\boldsymbol{t\to j\,\gamma\gamma}$}{t > j a a}}
\vspace*{-2mm}

Finally, the $C_{u\varphi}$ coefficients can be bound using the $t\to
j\gamma\gamma$ search presented in Ref.~\cite{CMS:2014qxa}. As mentioned
before, the interpretation the CMS Collaboration overlooks a
dependence in the flavor-changing $t\gamma q$ couplings. Imposing
$m_{\gamma\gamma}\in[120,130]$\,GeV and $m_{\gamma j}>10$\,GeV with
$m_h=125$~GeV and $\Br(h\to \gamma\gamma)=0.23\%$, we get:
\begin{align*}
\Gamma_{t\to j\gamma\gamma}^\text{on-$h$-peak}&=
	1.09\times10^{-6}\,\text{GeV}
	\quad (1\,\text{TeV}/\Lambda)^4\\
  \sum_{a=1,2}\Big\{
  	 &\big|C_{u\varphi}^{(a3)}|^2
  	+ \big|C_{u\varphi}^{(3a)}|^2\\[-1.5mm]
  	+0.37\big(
  		&\big|C_{uB}^{(a3)}+C_{uW}^{(a3)}\big|^2
  		+\big|C_{uB}^{(3a)}+C_{uW}^{(3a)}\big|^2
  	\big)\Big\}
\end{align*}
at leading order, with the interference between the $C_{uB}+C_{uW}$ and
the $C_{u\varphi}$ contributions neglected. However, given the bounds set previously
on $C_{uB}+C_{uW}$ and the relatively mild constraint on $t\to j\,\gamma\gamma$,
those $t\gamma q$ contributions have no significant impact on the global limits
we set. We therefore consider the following NLO~\cite{Zhang:2013xya} constraint instead:
\begin{align}
\label{eq:pj_bound}
  \sum_{a=1,2}\Big\{
  & \big|0.9997\:C_{u\varphi}^{(a3)}-0.0243\:C_{uG}^{(a3)}\big|^2\nonumber\\[-4mm]
 +& \big|0.9997\:C_{u\varphi}^{(3a)}-0.0243\:C_{uG}^{(3a)}\big|^2\Big\}\nonumber\\[-1mm]
  &\quad<12.8\;
  	\frac{0.69\%}{\Br_{t\to jh}^\text{exp}}\;
  	\frac{0.23\%}{\Br_{h\to \gamma\gamma}}\;
  	\frac{\Gamma_t}{1.32\,\text{GeV}}
  	\left(\frac{\Lambda}{1\,\text{TeV}}\right)^4.
 \nonumber
\end{align}

\begin{figure}[tb]
\includegraphics[scale=1]{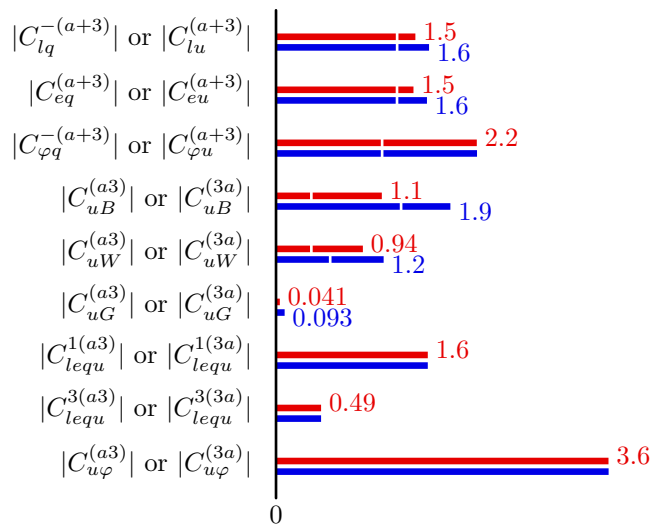}
\caption{The $95\%$ CL limits on the moduli of operator coefficients for
$\Lambda =1\,\text{TeV}$, as deriving from current bounds on $t\to
j\,\ell^+\ell^-$, $t\to j\gamma\gamma$, $pp\to t\!+\!\bar t$, $pp\to t\gamma\!+\!\bar
t\gamma$ and $e^+e^-\to tj\!+\!\bar tj$. Two-quark--two-lepton operators containing
an electron pair are shown, for the one containing a muon pair we refer to the
figures in \autoref{sec:gluon} and \autoref{sec:photon}. The red allowed regions
apply for $a=1$ and the blue ones for $a=2$. A white mark indicates
the bound that would have been obtained by fixing all coefficients to zero but
the one constrained, instead of performing a global analysis.
}
\label{fig:limits}
\end{figure}

Unfortunately, a statistical combination of the $95\%$ CL bounds derived in
this section is not possible with the
published information. We can only require those constraints to be
simultaneously satisfied. The results of this procedure are shown in
\autoref{fig:limits}.

\vspace*{-3mm}
\section{Conclusions}\label{sec:conclusions}
\vspace*{-2mm}

A fully gauge-invariant effective field theory allows the consistent, global and
accurate interpretation of new-physics searches in terms of well-defined theoretical
parameters. Our global analysis at NLO in QCD of the most constraining limits on
top-quark FCNC operators provides a proof-of-principle of feasibility of this program.

In particular, we have stressed the importance of considering simultaneously all
contributions arising at dimension six in the standard-model effective theory,
four-fermion operators included. Separating on-$Z$-peak and off-$Z$-peak lepton
invariant mass regions in $t\to j\,\ell^+\ell^-$ searches would allow us to better
constrain two-quark--two-lepton operators, especially the ones involving a muon
pair. Distinguishing the lepton channels would permit us to bound accurately
different operators. In general, efficiencies for each contribution and fiducial
limits should be made public. Angular distributions ---helicity fractions
notably--- would provide additional separation power between operators of
different Lorentz structures. The effort devoted to the searches of top-quark
FCNC production processes should be pursued further as they probe higher energy
scales than decays. In particular, an update of the limit on $pp\to
t\;\ell^-\ell^+$ and a search for $pp\to th$ would probably improve
significantly the constraints presented here. Ultimately, the publication of the
statistical information from which the limits derive would allow for a more
appropriate combination of the constraints coming from different observables.

In this work, we have moved the first steps towards a global approach to the
determination of the top-quark couplings in the context of an effective field
theory by considering the case of FCNC interactions at NLO in QCD. The same
approach can be extended to flavor-conserving and charged-current interactions.
The impact of indirect constraints arising from $B$ mesons, electroweak or Higgs
data could (and should) also be considered. In this respect the effective
field theory provides a unique framework where all information coming from
different measurements and observables can be consistently, accurately and
precisely combined to set bounds on new physics.

\vspace*{-4mm}
\section*{Acknowledgements}
\vspace*{-2mm}
We would like to thank Mojtaba Najafabadi, Reza Goldouzian and Andrea Giammanco
for details about the analysis of Ref.~\cite{CMS:2014hwa}. This work has been
performed in the framework of the ERC grant 291377 ``LHCTheory'' and of the FP7
Marie Curie Initial Training Network MCnetITN (PITN-GA-2012-315877). C.~Z.~has
been supported by the IISN ``Fundamental interactions'' convention 4.4517.08,
and by US Department of Energy under Grant DE-AC02-98CH10886. G.~D.~is a Research
Fellow of the FNRS, Belgium, and of the Belgian American Education Foundation,
USA.

\newpage
\onecolumngrid
\appendix
\section{NLO corrections to \texorpdfstring{$\boldsymbol{e^+e^-\to tj}$}{e e > t j}}\label{sec:k-fac}
NLO cross sections for $e^+e^-\to tj$ can be written as
\begin{equation*}
  \sigma^{\text{NLO}}=\sigma^{\text{LO}}\left[ 1+\frac{\alpha_s(\mu)}{\pi}\delta(x,\mu) \right]
\end{equation*}
where $\mu$ is the renormalization scale, and $x=m_t/\sqrt{s}$. The quantum
corrections $\delta$ depend on the Lorentz structure of the quark current.We obtain, for a vector current,
\begin{align*}
\delta(x,\mu)=&\frac{1}{3 \left(1-x^2\right) \left(2+x^2\right)}
\Bigg[
  6 -9 x^2 -5 x^4 +\frac{4 x^2}{1-x^2} \left(5 x^4-4 x^2-5\right) \log (x)
\nonumber\\
&-2 \left(1-x^2\right) \left(5 x^2+4\right) \log \left(1-x^2\right)
+8 \left(1-x^2\right)\left(2+x^2\right) \left(\log (x) \log \left(1-x^2\right)
+\text{Li}_2\left(x^2\right)\right)
\Bigg]\ ,
\end{align*}
for a scalar current,
\begin{align*}
\delta(x,\mu)=& \frac{1}{3}\Bigg[
-6 \log \left(\frac{s}{\mu^2}\right)
+17 +8 \frac{x^2 \left(x^2-2\right)}{\left(1-x^2\right)} \log (x)
+2 \left(2x^2-5\right) \log \left(1-x^2\right)
+8 \log (x) \log \left(1-x^2\right)
+8 \text{Li}_2\left(x^2\right)
\Bigg]\ ,
\end{align*}
and, finally, for a tensor current,
\begin{align*}
\delta(x,\mu)=&\frac{1}{9 \left(1-x^2\right) \left(1+2 x^2\right)}
  \Bigg[
    6 \left(1-x^2\right)(1+2x^2)\left( \log \left(\frac{s}{\mu^2}\right)
    +4\log (x) \log \left(1-x^2\right)+4\text{Li}_2\left(x^2\right)\right)
\nonumber\\&
  +24 \frac{5 x^6-7 x^4+x^2}{1-x^2} \log (x)
    -6 \left(1-x^2\right) \left(1+8 x^2\right) \log \left(1-x^2\right)
    -32x^4+13x^2+7
    \Bigg]\ .
\end{align*}
\twocolumngrid

\bibliographystyle{apsrev4-1_title}
\bibliography{FCNC_NLO}
\end{document}